\documentclass[]{pasj01}
\draft

\Received{2020/06/26}
\Accepted{2020/08/26}
 
\begin{document} 

\title{
GPU-accelerated Image Reduction Pipeline }

\author{
    Masafumi \textsc{Niwano}\altaffilmark{1},
    Katsuhiro L. \textsc{Murata}\altaffilmark{1},
    Ryo \textsc{Adachi}\altaffilmark{1},
    Sili \textsc{Wang}\altaffilmark{1},
    Yutaro \textsc{Tachibana}\altaffilmark{1},
    Youichi \textsc{Yatsu}\altaffilmark{1},
    Nobuyuki \textsc{Kawai}\altaffilmark{1}%
}
\author{Takashi \textsc{Shimokawabe}\altaffilmark{2}}

\author{Ryousuke \textsc{Itoh}\altaffilmark{1,3}}

\email{niwano@hp.phys.titech.ac.jp}

\altaffiltext{1}{Department of Physics, Tokyo Institute of Technology, 2-12-1 Ookayama, Meguro-ku, Tokyo 152-8551, Japan}
\altaffiltext{2}{The Supercomputing Division, Information Technology Center, The University of Tokyo, 2-11-16 Yayoi, Bunkyo-ku, Tokyo 113-8658, Japan}
\altaffiltext{3}{Bisei Astronomical Observatory, 1723-70 Ookura, Bisei-cho, Ibara, Okayama 714-1411, Japan}


\maketitle

\begin{abstract}
We developed a high-speed image reduction pipeline using Graphics Processing Units (GPUs) as hardware accelerators.
Astronomers desire detecting EM counterpart of gravitational-wave sources as soon as possible for sharing positional information to organize systematic follow-up observations.
Therefore, high-speed image processing is important.
We developed a new image reduction pipeline for our robotic telescope system, which uses a GPU via a Python package CuPy to achieve high-speed image processing.
As a result, the processing speed was increased by more than a factor of forty to that of the current pipeline, while maintaining the same functions.
\end{abstract}

\section{Introduction}\label{sec:intro}
Gamma-ray bursts (GRB) and gravitational waves (GW) are short-lived ($\lesssim10^{2-3}$sec) and no one knows when or where they occur, and we call such phenomena as ``transients''.
In order to understand the physical processes of transients, it is essential to perform multi-wavelength and continuous observations for before, during, and for a reasonable length of time after the occurrence.
We aim for observing their afterglow and perform the prompt follow-up observations.
However, positional information about the GW source obtained from laser interferometers such as LIGO \citep{ligo1}, Virgo \citep{virgo1} and KAGRA \citep{kagra1} has too large of an uncertainty ($\approx10^{1-3}\rm deg^2$) to perform systematic follow-up observations.
Thus it is necessary to immediately identify the EM counterpart and send more accurate information to the ground telescope network as soon as possible.
To realize that, all processes from receiving an alert to finding the counterpart must be performed in a short period ($\lesssim0.1$days), and therefore, high-speed image processing becomes the key technology.

High-speed image processing is especially important in the era of time-domain astronomy which requires high data-rate observations consisting of a large number of pixels taken at high cadence.
Wide-field surveys such as Pan-STARRS (\cite{ps1}), Subaru-HSC (\cite{hsc1}), ZTF (\cite{ztf1}), and Tomo-e Gozen \citep{tomo-e1} are typical examples.
One of the solutions to deal with the high data-rate is CPU-based parallel computing.
For example, ZTF obtains very high-resolution CCD images of about 600 Mpix every 40 sec, and its data rate exceeds 200 Mbit/s.
To process those extremely fast-growing data, they perform parallel computing in a computing system with 1192 physical CPU cores. As a result, they have achieved the real-time detection of transients and moving objects \citep{ztf2}.

A Graphics Processing Unit (GPU) is a hardware-accelerator for graphics processing.
A GPU has thousands of computing units called ``shaders'', which correspond to CPU cores, and the GPU achieves extremely high performance only for SIMD\footnote{Single Instruction Multiple Data} processing (e.g. matrix operation).
Although that is achieved by specializing in graphics processing, this performance is sometimes utilized for not only graphics processing, but also other purposes, and such GPU utilization is called GPGPU (General Purpose computing on GPU).
In the astronomy, GPGPU is applied for physical simulation, machine learning, analysis of radio interferometer data, to name of few (\cite{gpu1}, \cite{gpu2}, \cite{gpu3}).
To build the same-scale computing system as ZTF, 120 Xeon E5-2640v4\footnote{CPU used at ZTF Science Data System} processors are required and these consume about 10 kW of power.
However, with Tesla P100, the same theoretical performance can be achieved with only 10 modules, and the power consumption will be less than 2.5 kW (table \ref{tab:xeon&tesla}).
That advantage is limited to the case when the performance of the GPU can be fully exploited, but it is not so difficult because image processing is a speciality of the GPU.

\begin{table}
  \tbl{
    Performance comparison of CPU and GPU}{
    \begin{tabular}{llrr}\hline
    & Name & FP64\footnotemark[1] & Power\footnotemark[2] \\
    & & [TFLOPS] & [W]\\\hline
    CPU & Intel Xeon E5-2640v4 & 0.4 & 90\\
    GPU & NVIDIA Tesla P100 (PCIe) & 4.7 & 250\\\hline
    \end{tabular}
    }\label{tab:xeon&tesla}
\begin{tabnote}
\footnotemark[1] Theoretical maximum number of 64-bit floating-point operations performed per second. The CPU value is calculated with IPC=16 (instructions per clock) and Clock=2.4GHz, and the GPU value is nominal.\\
\footnotemark[2] The CPU value is the thermal design power, and the GPU value is the maximum power consumption.\\
\end{tabnote}
\end{table}

The purpose of this research is to utilize GPU for astronomical image processing, and more specifically, to speed up the image reduction pipeline of our robotic telescopes, MITSuME-Akeno and MITSuME-Okayama (\cite{mtm1}, \cite{mtm2}, \cite{mtm3}).


\section{Development}\label{sec:develop}
\subsection{Image reduction at MITSuME}\label{sec:oldpipeline}
Both of MITSuME telescopes are equipped with three CCD cameras to obtain simultaneously three-band images.
Each CCD generates a 16-bit integer image of 1074$\times$1024 pixels including an overscan area, which is stored as a FITS file.
Typical value of exposure time, overhead and data volume is 60 sec (for GW or GRB), 8 sec and 4 GB/night (for 2 telescopes), respectively.
To reduce the pixel randomization, we perform 9 points dithering.
Therefore positions of point sources on the detector differs for each exposure even observing the same object.
After derivation of pedestal level and astrometry, FITS files are transferred to the data server in Tokyo Tech.

\begin{figure}
 \begin{center}
  \includegraphics[width=\hsize]{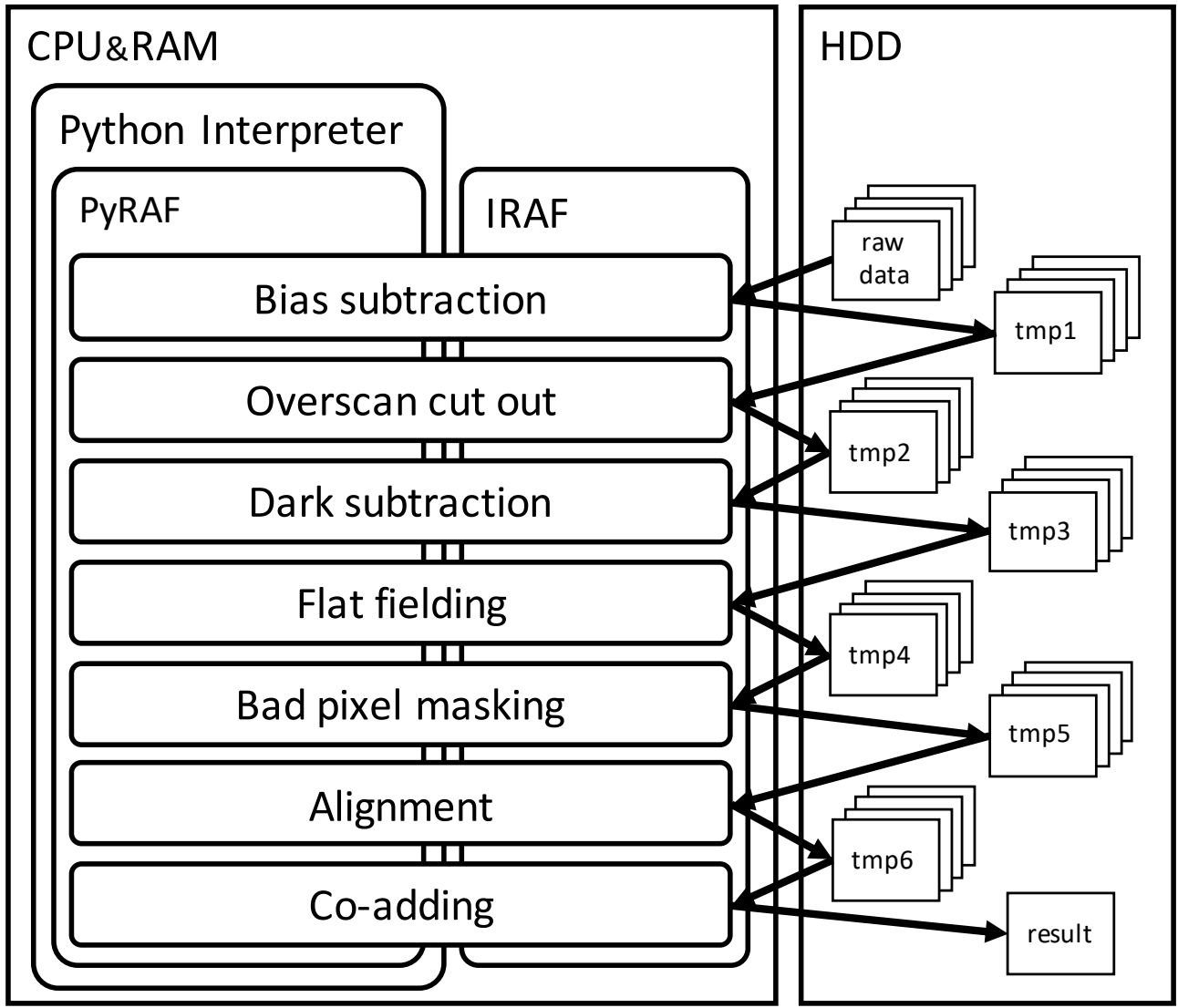} 
 \end{center}
\caption{Current Pipeline overview:
    Rectangular objects in a frame labeled ``HDD'' (Hard Disk Drive) indicate FITS files, ``raw data'' is a set of unreduced data, ``result'' is a reduced data, and ``tmp$*$'' are sets of a temporary file.
    Arrows indicate data flows.
    Only the processing performed by IRAF is shown, and there are actually other processing.
}
\label{fig:currentoverview}
\end{figure}

The current image reduction pipeline is implemented as a Python\footnote{https://www.python.org/} script which uses IRAF\footnote{http://ast.noao.edu/data/software} (\cite{iraf1}, \cite{iraf2}) via PyRAF\footnote{http://www.stsci.edu/institute/software\_hardware/pyraf}, and its sequence is shown in figure \ref{fig:currentoverview}.
Specifically, overscan cutout, bias subtraction, dark subtraction, flat fielding, bad pixel masking, alignment, and co-adding are performed by calling IRAF tasks.
At the stage of bias subtraction, casting from 16-bit integer to 32-bit floating point is performed.
In the overscan cutout, overscans of $50\times1024$ pix and 2 pix from the edge are removed, leaving the image of $1020\times1020$ pix.
Because of the dithering, alignment is necessary before co-adding.
The pipeline uses bicubic-spline interpolation and sigma-clipped-mean algorithm for sub-pixel image shifting and co-adding, respectively.
Only the processing performed by IRAF/PyRAF is shown here, and other processing such as calculating the relative position of images using WCSTools\footnote{http://tdc-www.harvard.edu/wcstools/} is not shown.

\subsection{New Reduction Pipeline}\label{sec:newpipeline}
We created a new pipeline by using Python, because Python is an easy-to-code language, and there are Python packages available for the purpose of this research.
The new pipeline uses NumPy\footnote{https://numpy.org/} \citep{numpy2}, Astropy\footnote{https://www.astropy.org/index.html} \citep{astropy2}, PyWCS\footnote{https://pypi.org/project/pywcs/} and CuPy\footnote{https://cupy.chainer.org/}.
CuPy is a Python package for numerical calculation on GPU using CUDA\footnote{https://developer.nvidia.com/cuda-zone}.
The most remarkable feature of CuPy is the API compatibility with NumPy, thus a code written with NumPy can be easily rewritten with CuPy.

\begin{figure*}
 \begin{center}
  \includegraphics[width=0.7\hsize]{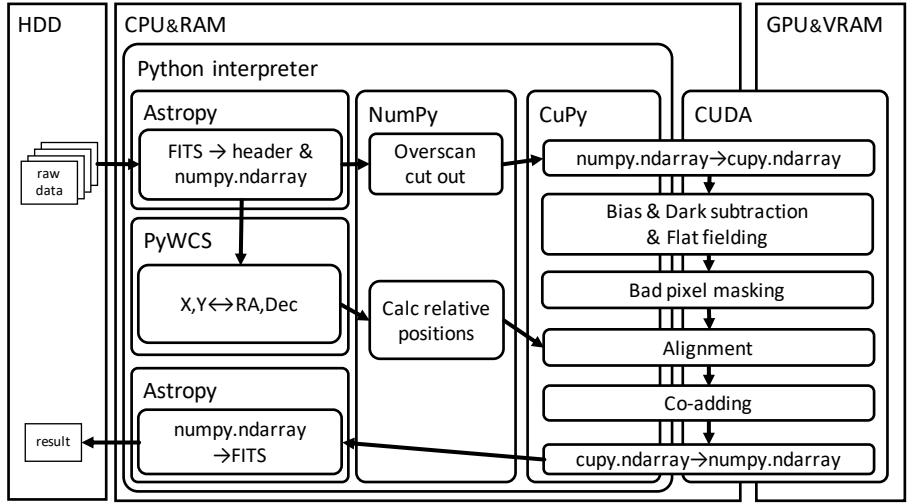} 
 \end{center}
\caption{
    New pipeline overview:
    Rectangular objects on ``HDD'' indicate FITS files, ``raw data'' is a set of unreduced data, ``result'' is a reduced data.
    Arrows indicate data flows.
    Roughly, ``raw data'' are read by Astropy, array calculation is performed by CuPy, and the result is written to ``result'' by Astropy.
}\label{fig:overview}
\end{figure*}

The sequence of the new pipeline is described in figure \ref{fig:overview}.
Specifically, the header and image data are read from FITS file as instances of \texttt{astropy.io.fits.Header}\footnote{Python class provided by Astropy for storing contents of FITS header} and \texttt{numpy.ndarray}\footnote{Python class of multidimensional array provided by NumPy}.
At the same time, casting from 16-bit integer to 32-bit floating point is also performed.
The header is passed to \texttt{pywcs.WCS}\footnote{Python class for converting coord between WCS and other system} constructor for calculating the relative position of the images, and the image data is converted to \texttt{cupy.ndarray}\footnote{Python class corresponding to \texttt{numpy.ndarray} in CuPy} while slicing overscan. 
Then, bias subtraction, dark subtraction, flat fielding, bad pixel masking, alignment, and co-adding are performed on the GPU with CuPy.
The array of resulting image is converted to \texttt{numpy.ndarray}, and then written to a FITS file with Astropy.
The pipeline uses bicubic-spline interpolation and sigma-clipped-mean algorithm, the same as the current pipeline.
The image processing is performed on the GPU-side because it is basically a SIMD operation, and the other processing is performed on the CPU-side because there are few benefits performing it on the GPU or it cannot be performed on the GPU.
The overscan cutout is performed on the latter because the array-slicing operation is not SIMD, and transferring data not used for calculation to the VRAM generates extra memory allocation time.
\texttt{astropy.wcs.WCS} has the same functions as \texttt{pywcs.WCS}, but an initialization of the former takes $\approx7\rm ms$ while the latter takes $\approx200\rm\mu s$ in our environment.

\section{Verification and Performance Test}\label{sec:performance}

We evaluated consistency and performance of the new pipeline.
Information about the execution environment is shown in tables \ref{tab:hardware} and \ref{tab:software}.

\begin{longtable}{lll}
  \caption{Machine Specification}\label{tab:hardware}
  \hline              
   & Model & Notes \\ 
\endfirsthead
  \hline
   & Model & Notes \\
\endhead
\endfoot
\endlastfoot
  \hline
    M/B & Supermicro X10SRA & C612\\
    CPU & Intel Xeon E5-1650 v4 & 3.6-4.0GHz\\
    RAM & A2ZEON RD4R16G44S2400 $\times$4 & DDR4 2400MHz Registered ECC, Quad Channel\\
    GPU & NVIDIA TITAN X (Pascal) & 3584 shaders, 1417-1531MHz, GDDR5X 12GB, PCIe 3.0 x16\\
    SSD & Intel SSDSC2BB12 & SATA 6Gbps, OS and executable files are located\\
    HDD & Seagate ST2000VN0001& SATA 6Gbps, Script files and data are located\\\hline
\end{longtable}
\begin{table}
  \tbl{Software versions}{
    \begin{tabular}{ll|ll}\hline
    OS&\multicolumn{3}{l}{Ubuntu 14.04 LTS x86\_64}\\\hline
    Python & 2.7.6 & CUDA & 10.0\\
    IRAF & 2.16.1 & NumPy & 1.15.4\\
    PyRAF & 2.1.14 & Astropy & 2.0.9\\
    SExtractor & 2.8.6 & PyWCS & 1.12\\
    WCSTools & 3.8.7 & CuPy & 5.1.0\\\hline
    \end{tabular}
    }\label{tab:software}
\begin{tabnote}
\footnotemark[$*$] Python packages other than PyRAF were installed with \texttt{pip}.  \\ 
\end{tabnote}
\end{table}

\subsection{Consistency Evaluation}\label{sec:consistency}
We coded the bicubic-spline algorithm and sigma-clipped-mean algorithm ourselves, because CuPy did not provide corresponding functions.
Therefore we had to confirm whether the same result as the current pipeline can be reproduced for these two processing methods.
We defined $\Delta\hat{I}$ as a quantity for evaluating relative difference between two images as follows,
\begin{equation}\label{eq:relative-difference}
    \Delta\hat{I}_{ij} \equiv \frac{I_{ij}-I^{\rm ref}_{ij}}{I^{\rm ref}_{ij}}
\end{equation}
where $I_{ij}$ is the pixel value of the test image, $I^{\rm ref}_{ij}$ is the pixel value of the reference image, and $i,j$ are indices of a pixel in the image.

\subsubsection{Shift Consistency}\label{sec:shift}
\begin{figure}
    \begin{center}
    \begin{tabular}{cc}
        {\scriptsize Image1} & {\scriptsize Image2} \\
        \includegraphics[width=0.4\hsize]{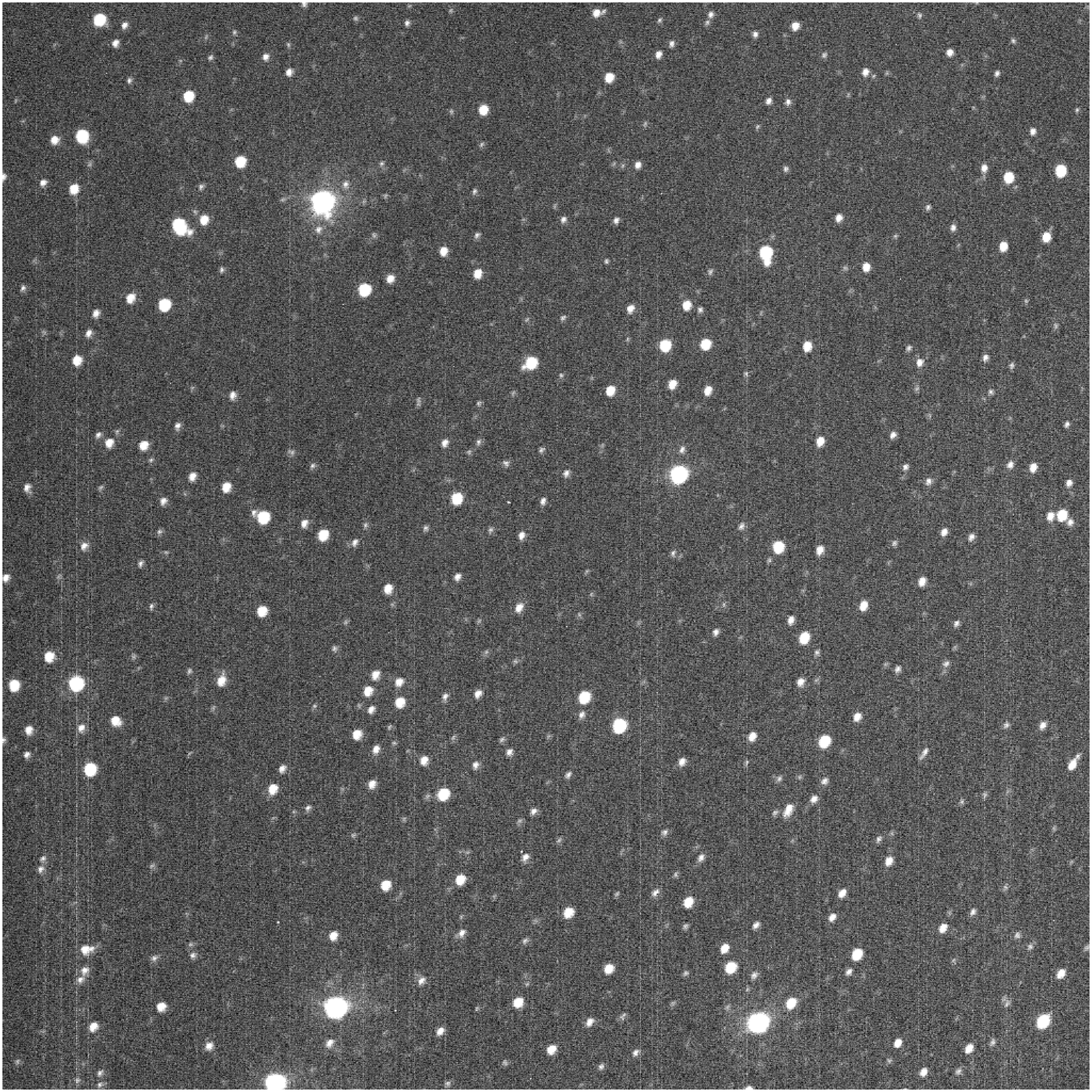} & \includegraphics[width=0.4\hsize]{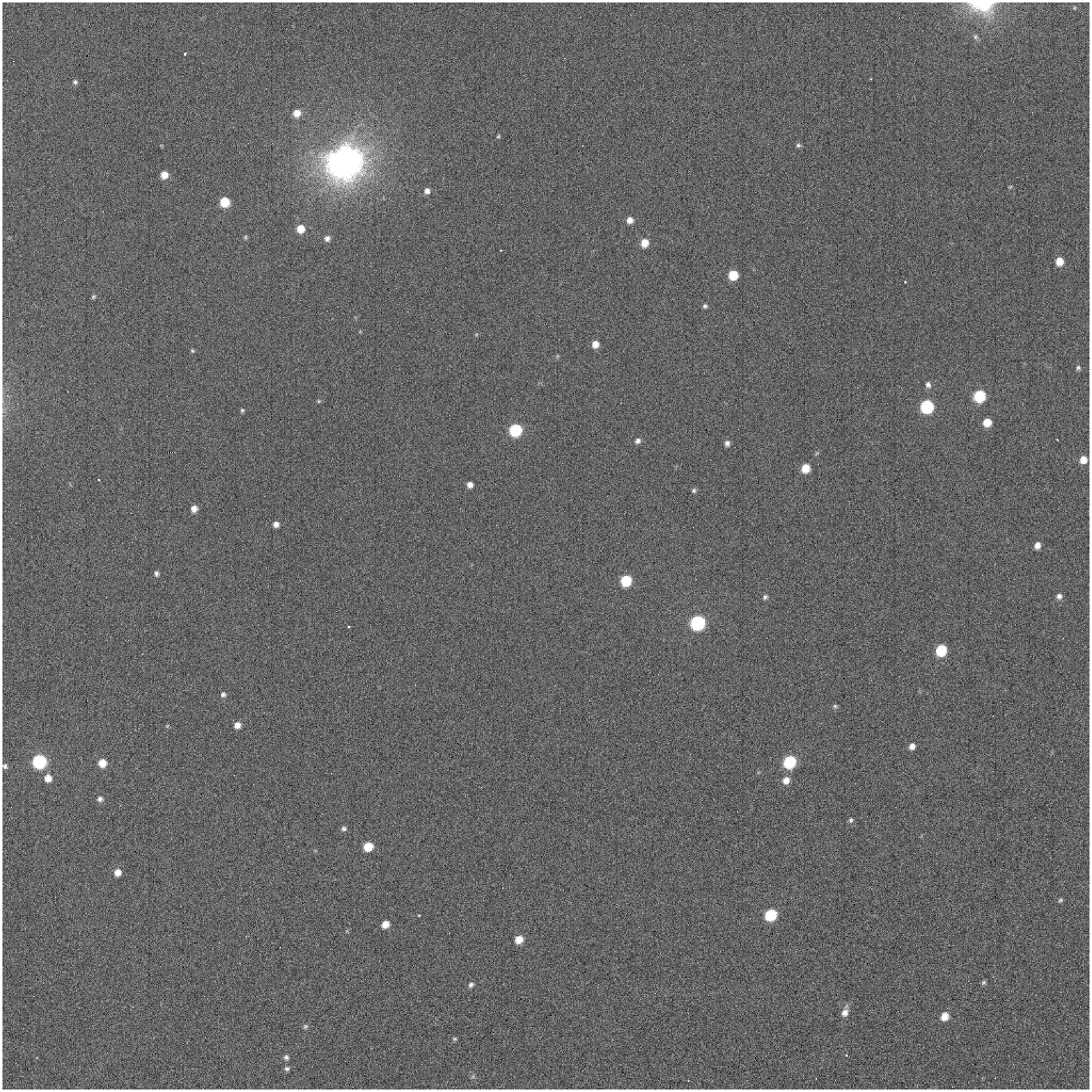} \\
        {\scriptsize Image3} & {\scriptsize Image4} \\
        \includegraphics[width=0.4\hsize]{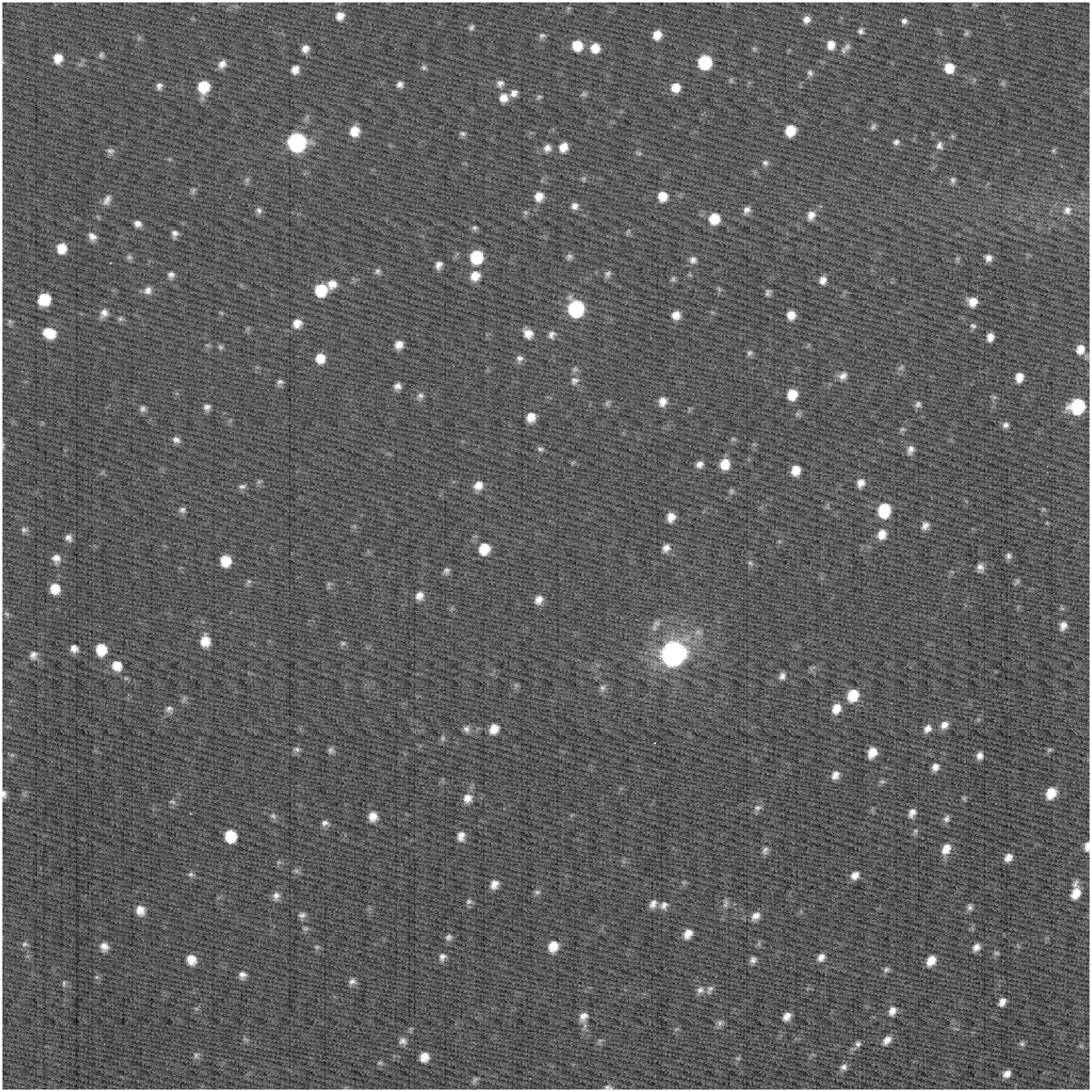} & \includegraphics[width=0.4\hsize]{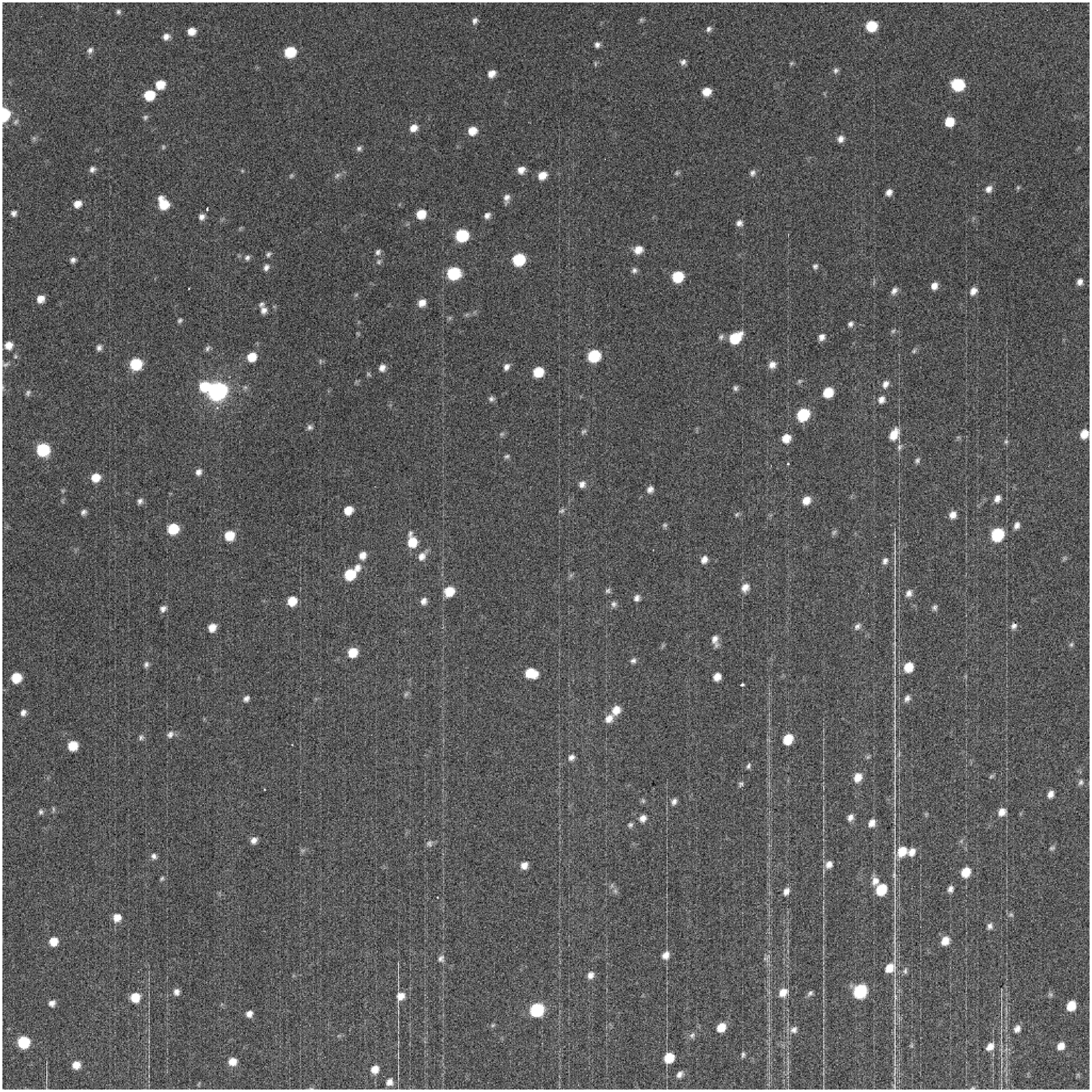}
    \end{tabular}
    \end{center}
    \caption{Data used for verification of shift consistency}
    \label{fig:shift}
\end{figure}

For the image processed up to bad-pixel masking in the reduction, we created the following two sets of images with various shift amounts, and calculated $\Delta\hat{I}$.
\begin{description}
    \item[A] The image shifted by the new pipeline, using a shifted image made by IRAF-\texttt{imshift} as a reference.
    \item[B] The image that was shifted once and then shifted back to the original position by the new pipeline, using the original image as a reference.
\end{description}
A is for evaluating the consistency with IRAF, and B is used to evaluating how much information of the original image is conserved. We defined a coordinate system in which the lower left corner of the image is the origin, the horizontal direction is the X-axis, and the vertical is the Y-axis.
Then, the shift amounts in the X, Y dimensions are $\rm\Delta X$, $\rm\Delta Y$.
The data used was obtained by applying overscan cut out, bias \& dark subtraction, flat fielding, and bad-pixel masking to four of the images acquired in MITSuME-Akeno on May 14, 2018. These are shown by the table \ref{tab:shift} and figure \ref{fig:shift}.
\begin{table}
  \tbl{Information of data used for verification of shift consistency}{
    \begin{tabular}{lrr}\hline
    Label\footnotemark[1] & BG-LEVEL\footnotemark[2] [ADU] & BG-RMS\footnotemark[3] [ADU]\\\hline
    Image1 & 705 & 23.741\\
    Image2 & 197 & 14.072\\
    Image3 & 68 & 13.514\\
    Image4 & 699 & 22.267\\\hline
    \end{tabular}
    }\label{tab:shift}
\begin{tabnote}
\footnotemark[1] An appropriate name to identify the data\\
\footnotemark[2] Median of pixel values\\
\footnotemark[3] Standard deviation of pixel values\\
\footnotemark[$*$] For all images, the shape is $1020\times1020$ and the numerical format is a 32-bit float.
\end{tabnote}
\end{table}

\begin{figure}
    \begin{center}
        \includegraphics[width=\hsize]{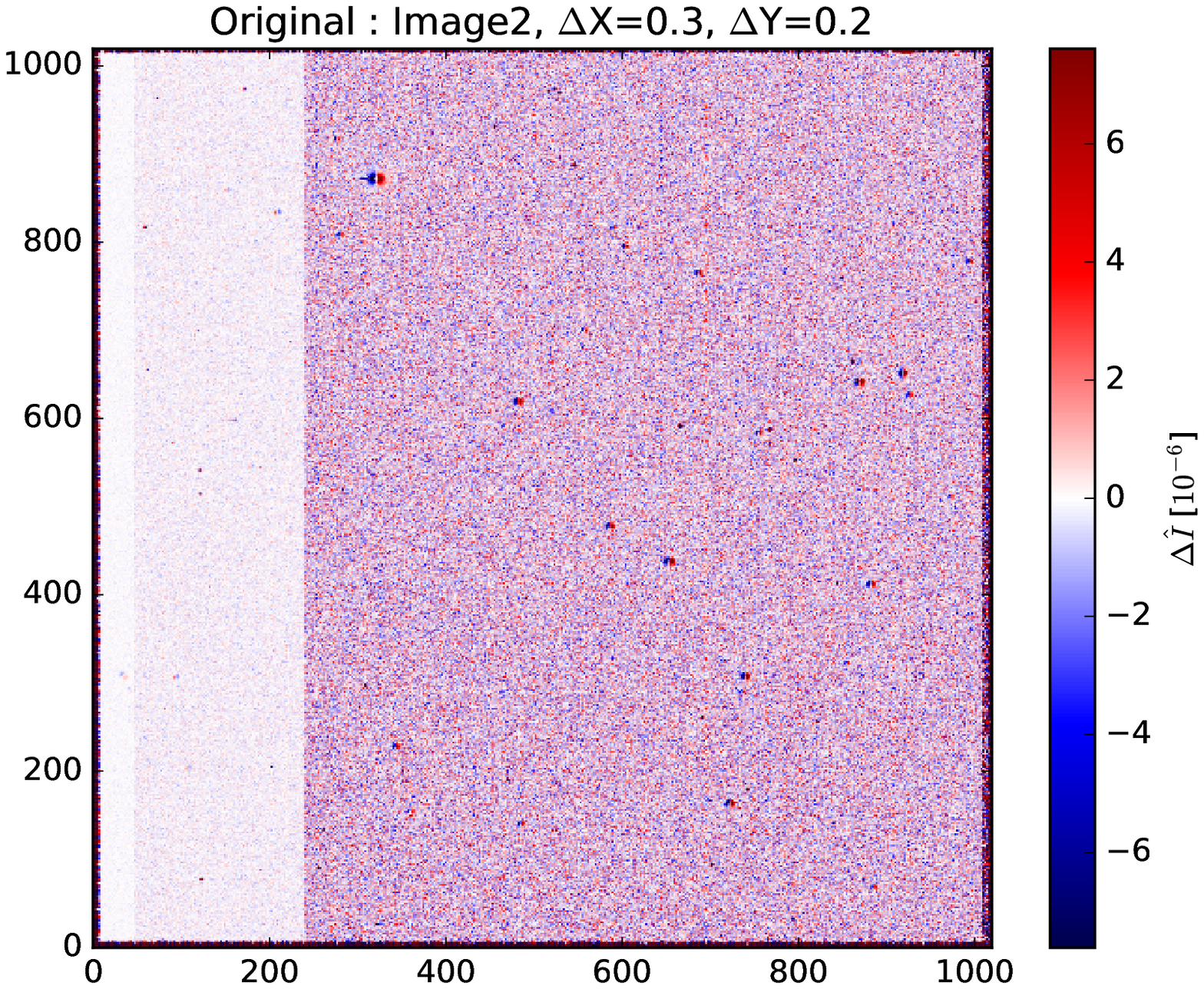}
    \end{center}
    \caption{An image in which each pixel is appropriately colored with $\Delta\hat{I}$ calculated for set A. Image2 is used, and $\rm(\Delta X,\Delta Y) = (0.3,0.2)$. 
    With the boundary $\rm X\approx240$, $|\Delta\hat{I}|$ is small on the left side and large on the right side.
    Around the locations where the original image has point sources, $\Delta\hat{I}$ is negative at near side from the boundary and positive at far side.
    }
    \label{fig:show_iraf_shift}
\end{figure}

\begin{figure}
    \begin{center}
        \includegraphics[width=\hsize]{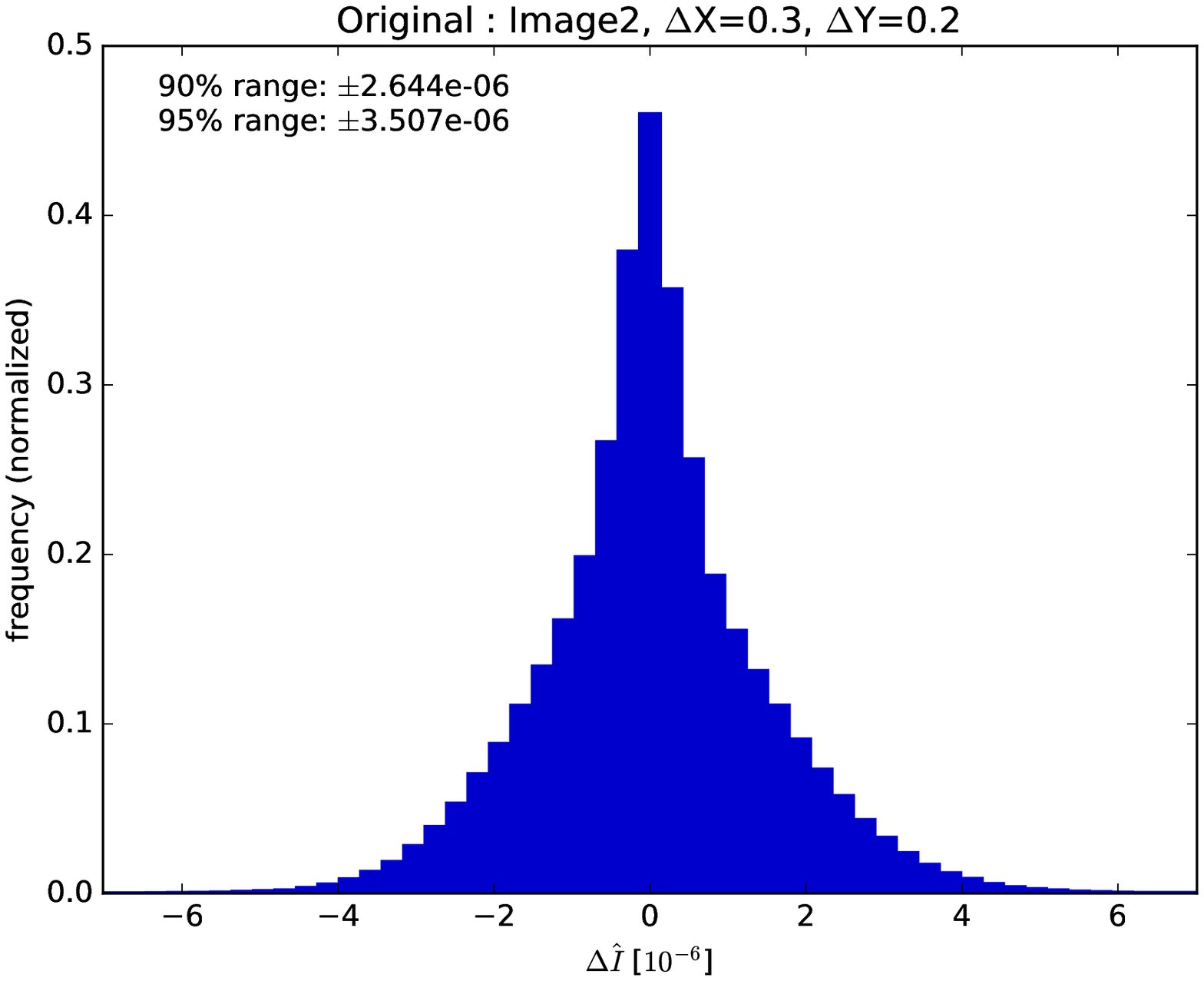}
    \end{center}
    \caption{A histogram of $\Delta\hat{I}$ calculated for set A. Image2 is used, and  $\rm(\Delta X,\Delta Y) = (0.3,0.2)$. ``90(95)\% range'' is the range in which 90(95)\% of $\Delta\hat{I}$ of all pixels fall. $\Delta\hat{I}$ is about $10^{-6}-10^{-5}$.}\label{fig:hist_iraf_shift}
\end{figure}

\begin{figure}
    \begin{center}
        \includegraphics[width=\hsize]{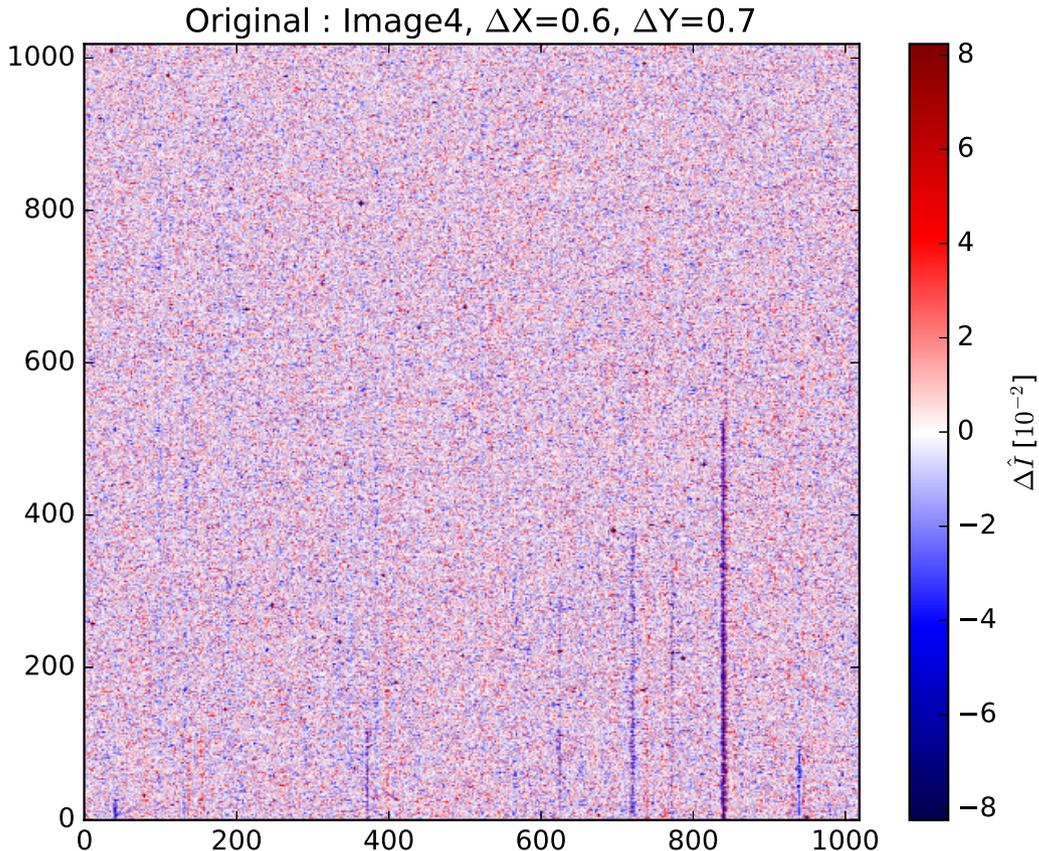}   
    \end{center}
    \caption{An image in which each pixel is appropriately colored with $\Delta\hat{I}$ calculated for set B . Image4 is used, and $\rm(\Delta X,\Delta Y) = (0.6,0.7)$. No particular pattern appears other than the pattern around the vertical line in the original image.}\label{fig:show_org_shift}
\end{figure}

\begin{figure}
    \begin{center}
        \includegraphics[width=\hsize]{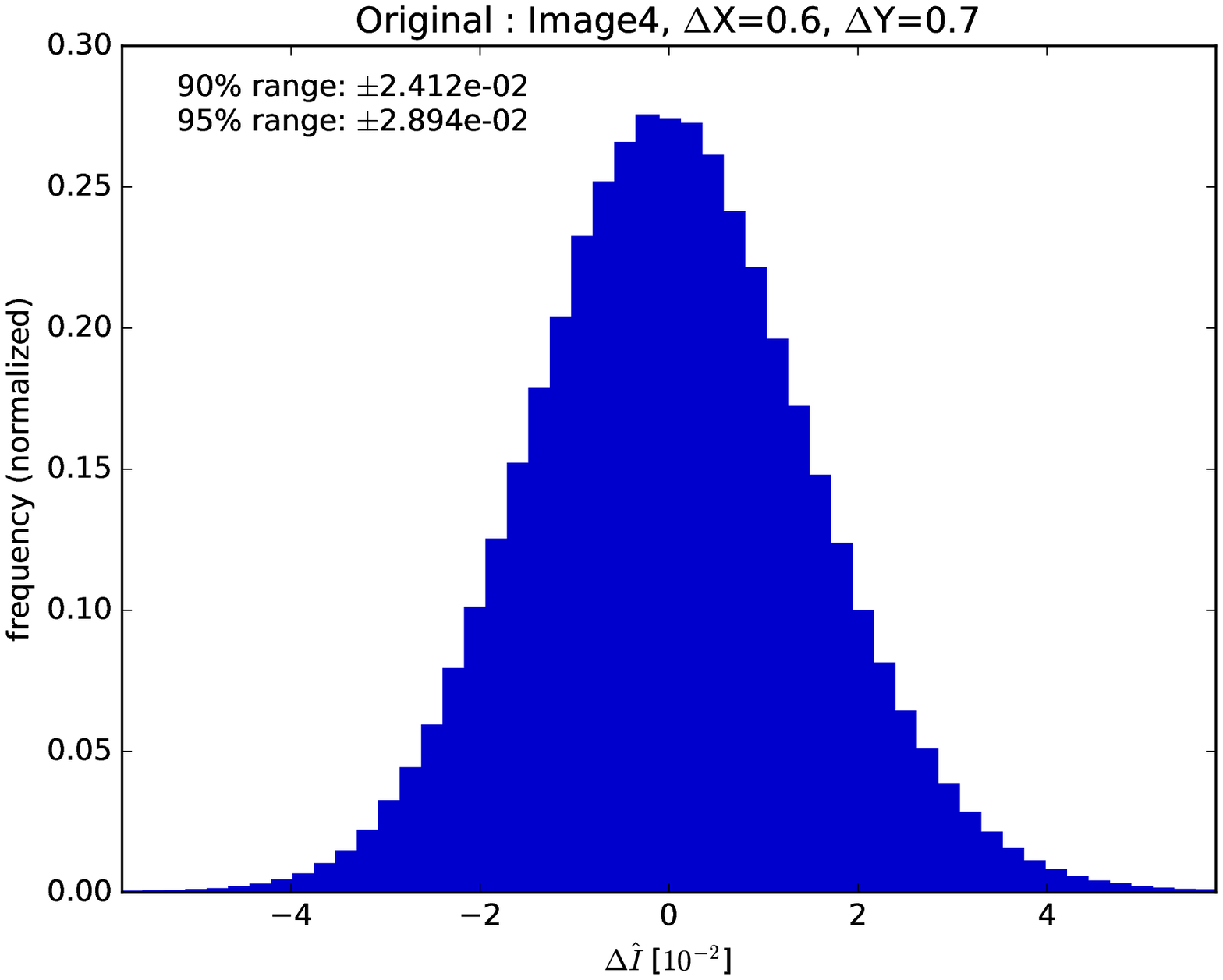}        
    \end{center}
    \caption{A histogram of $\Delta\hat{I}$ calculated for set B. Image4 is used, and  $\rm(\Delta X,\Delta Y) = (0.6,0.7)$. ``90(95)\% range'' is the range in which 90(95)\% of $\Delta\hat{I}$ of all pixels fall. $\Delta\hat{I}$ is about $10^{-2}-10^{-1}$.}\label{fig:hist_org_shift}
\end{figure}

Typical results are shown in figure \ref{fig:show_iraf_shift} and figure \ref{fig:hist_iraf_shift} for A, and figure \ref{fig:show_org_shift} and figure \ref{fig:hist_org_shift} for B.
Figure \ref{fig:show_iraf_shift} and figure \ref{fig:show_org_shift} are images in which each pixel is appropriately colored with $\Delta\hat{I}$, and figure \ref{fig:hist_iraf_shift} and figure \ref{fig:hist_org_shift} are histograms of $\Delta\hat{I}$.
The difference from IRAF is about $10^{-6}-10^{-5}$.
On the other hand, the strange pattern seen in figure \ref{fig:show_iraf_shift} suggests that this difference is not random but systematic.
Specifically, with the boundary $\rm X\approx240$, $|\Delta\hat{I}|$ is small on the left side and large on the right side.
Around the locations where the original image has point sources, $\Delta\hat{I}$ is negative at near side from the boundary and positive at far side.
The difference from the original image is about $10^{-2}-10^{-1}$, and no particular pattern as seen in figure \ref{fig:show_iraf_shift} appears.
The difference from IRAF is 3 to 4 orders of magnitude smaller than that from the original image.
Therefore, the effect of the shift operation itself on the result is more dominant than the effect of the change from IRAF to the new pipeline.

\begin{figure}
    \begin{center}
        \includegraphics[width=\hsize]{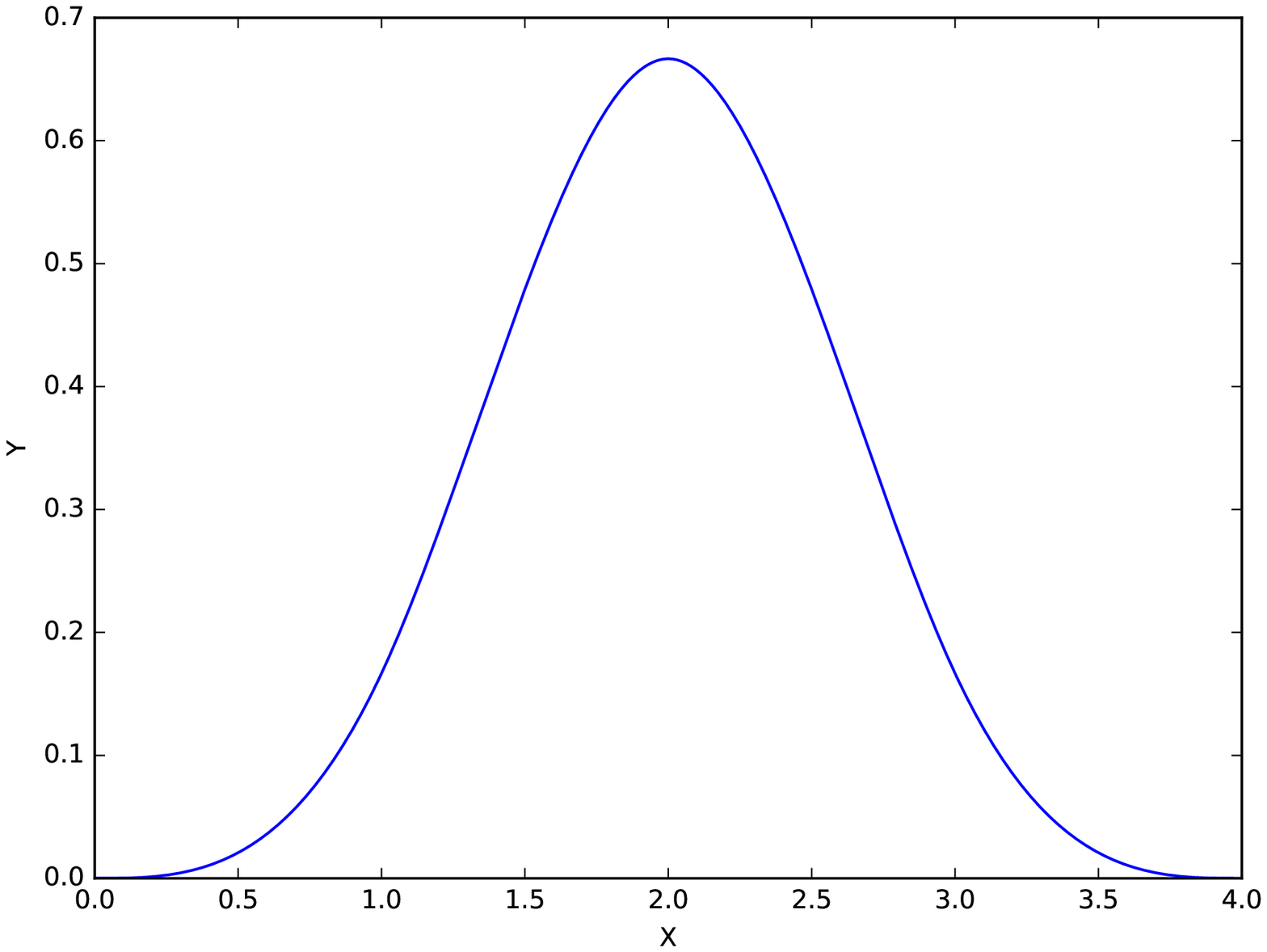}
    \end{center}
    \caption{Cubic B-spline basis function. This function satisfies the cubic-spline constraint conditions, so it can be reproduced exactly with cubic-spline.
    }
    \label{fig:bspline_graph}
\end{figure}

\begin{figure}
    \begin{center}
        \includegraphics[width=0.6\hsize]{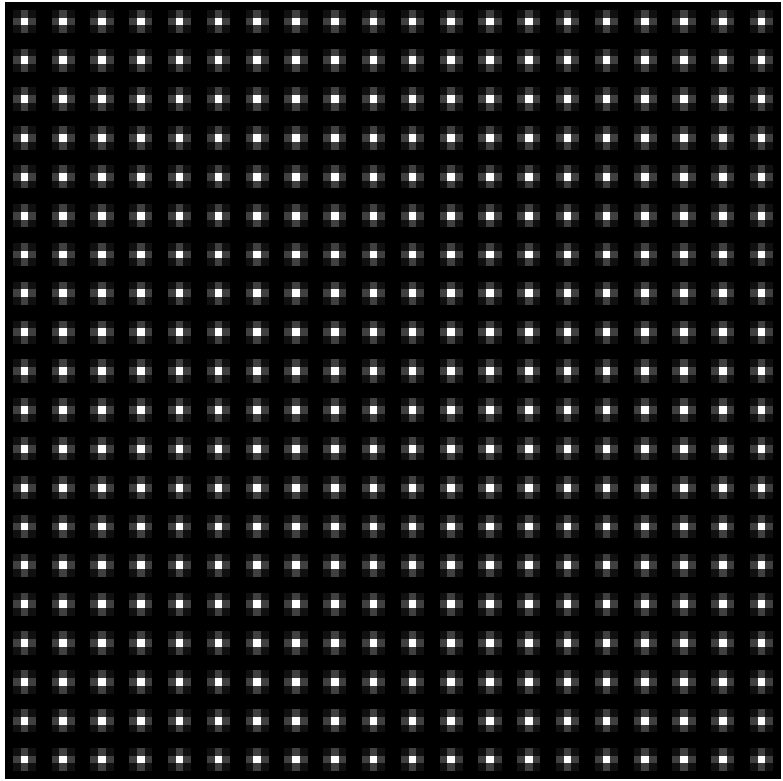}
    \end{center}
    \caption{Two-dimensional cubic B-spline basis functions were arranged in this image. A shifted image of this can be obtained without actually interpolating.
    }
    \label{fig:bspline_image}
\end{figure}

\begin{figure}
    \begin{center}
        \includegraphics[width=\hsize]{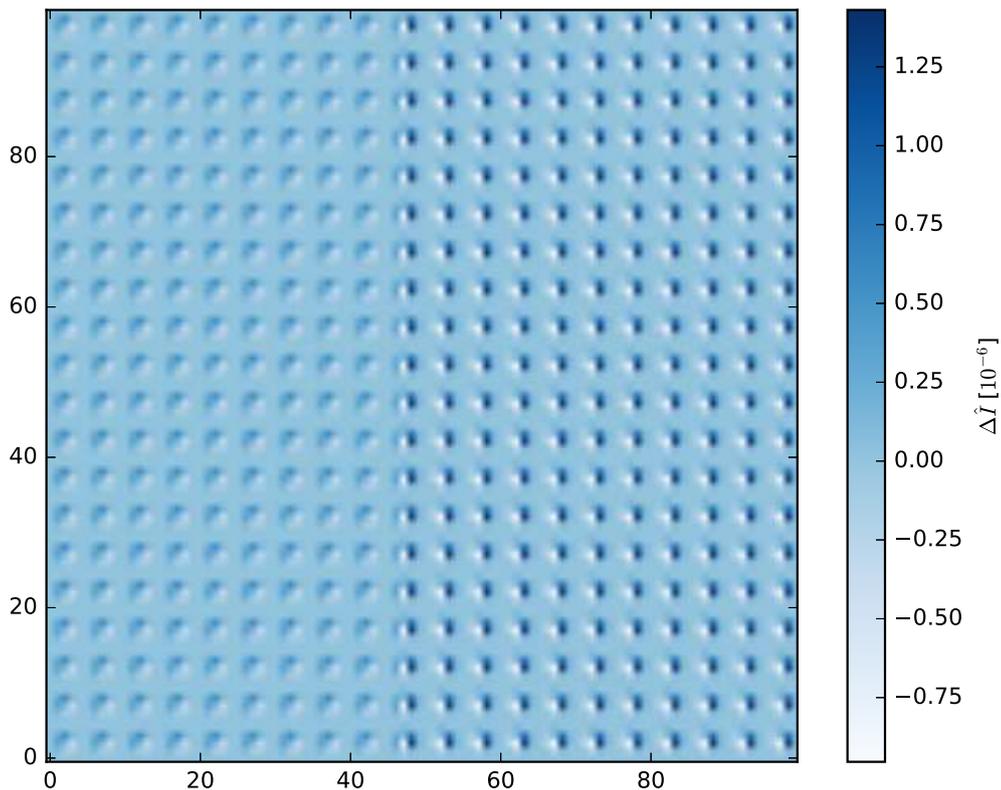}
    \end{center}
    \caption{An image in which each pixel is appropriately colored with $\Delta\hat{I}$ calculated for IRAF shifted image.
    $\rm(\Delta X, \Delta Y) = (0.3, 0.2)$.
    $\Delta\hat{I}$ is about $10^{-6}$ and the line of discontinuity as seen in figure \ref{fig:show_iraf_shift} appears.
    }
    \label{fig:diff_bsp_iraf}
\end{figure}

\begin{figure}
    \begin{center}
        \includegraphics[width=\hsize]{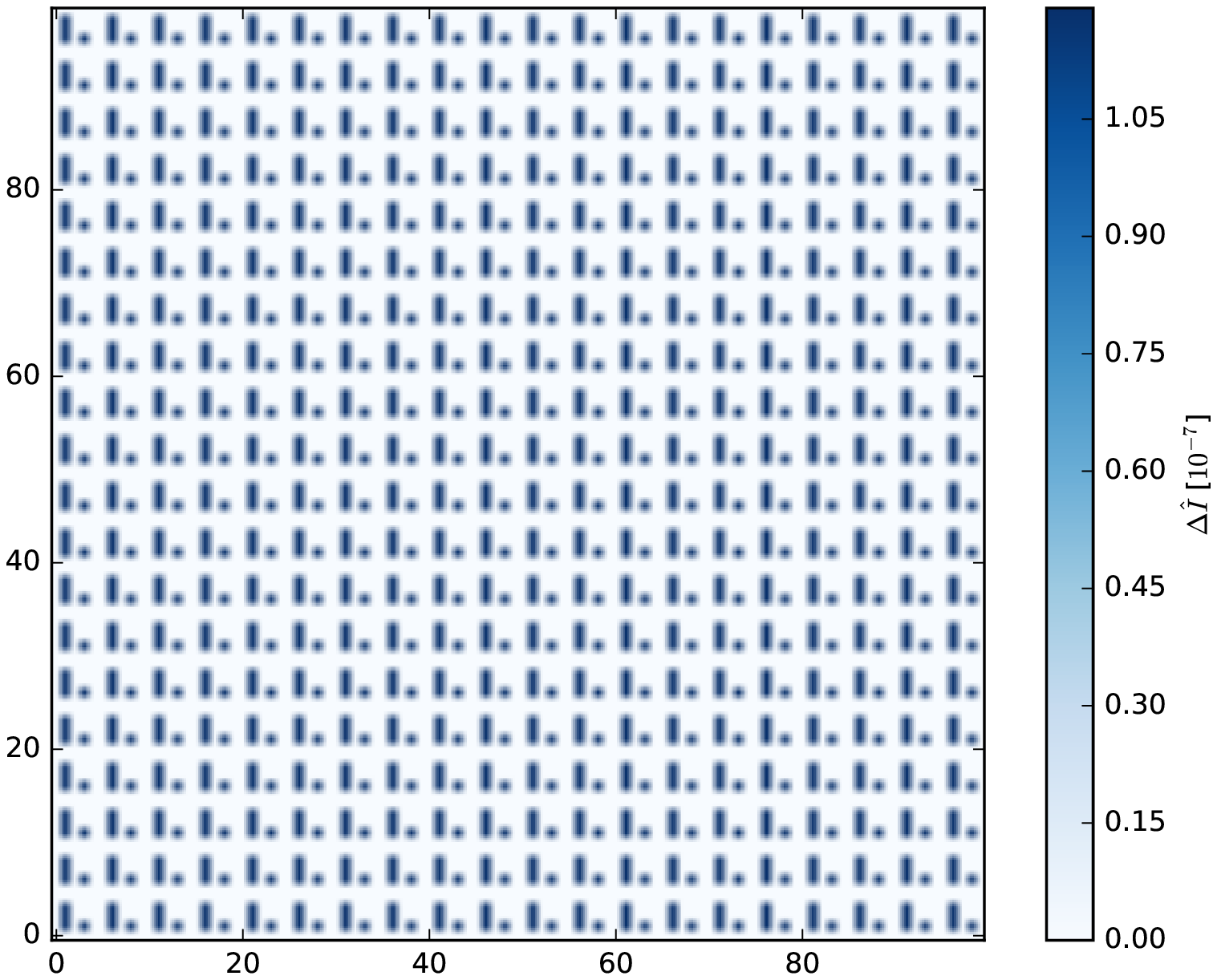}
    \end{center}
    \caption{An image in which each pixel is appropriately colored with $\Delta\hat{I}$ calculated for new pipeline shifted image.
    $\rm(\Delta X, \Delta Y) = (0.3, 0.2)$.
    $\Delta\hat{I}$ is about $10^{-7}$ and there is no discontinuity.
    }
    \label{fig:diff_bsp_new}
\end{figure}

We conducted additional experiments to determine the origin of the $10^{-6}-10^{-5}$ difference. 
First, we created an image in which two-dimensional cubic B-spline basis functions were arranged (figure \ref{fig:bspline_graph} and figure \ref{fig:bspline_image}).
Since the cubic B-spline basis function is exactly reproducible by cubic-spline interpolation, a shifted image can be obtained without actually interpolating.
Using the shifted image created by this way as a reference image, we obtained $\Delta\hat{I}$ for the image shifted by IRAF-\texttt{imshift} and the image shifted by new pipeline.
The results are shown in figure \ref{fig:diff_bsp_iraf} and \ref{fig:diff_bsp_new}.
For IRAF-\texttt{imshift}, $\Delta\hat{I}\approx10^{-6}$, and a line of discontinuity as shown in figure \ref{fig:show_iraf_shift} appears.
On the other hand, for the new pipeline, $\Delta\hat{I}\approx10^{-7}$, and there is no discontinuity.
Therefore, IRAF-\texttt{imshift} has a difference of about $10^{-6}$ from the accurate bicubic-spline interpolation.
(It has not been tested in other environments, and it is possible that the bug occurs only in our environment.)


\subsubsection{Co-adding Consistency}\label{sec:combine}
\begin{figure}
    \begin{center}
    \begin{tabular}{cc}
        {\scriptsize Set1} & {\scriptsize Set2} \\
        \includegraphics[width=0.4\hsize]{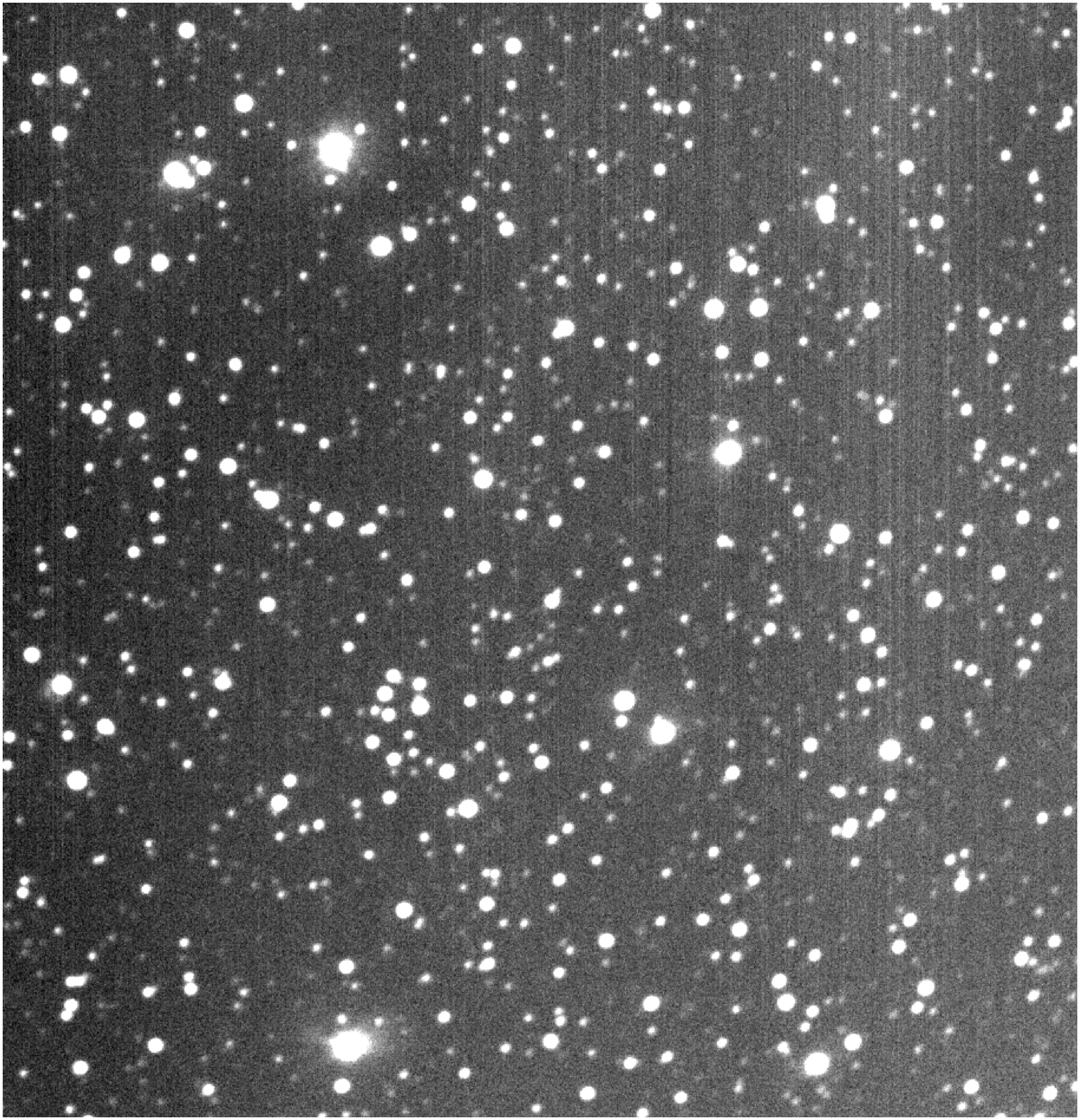} & \includegraphics[width=0.4\hsize]{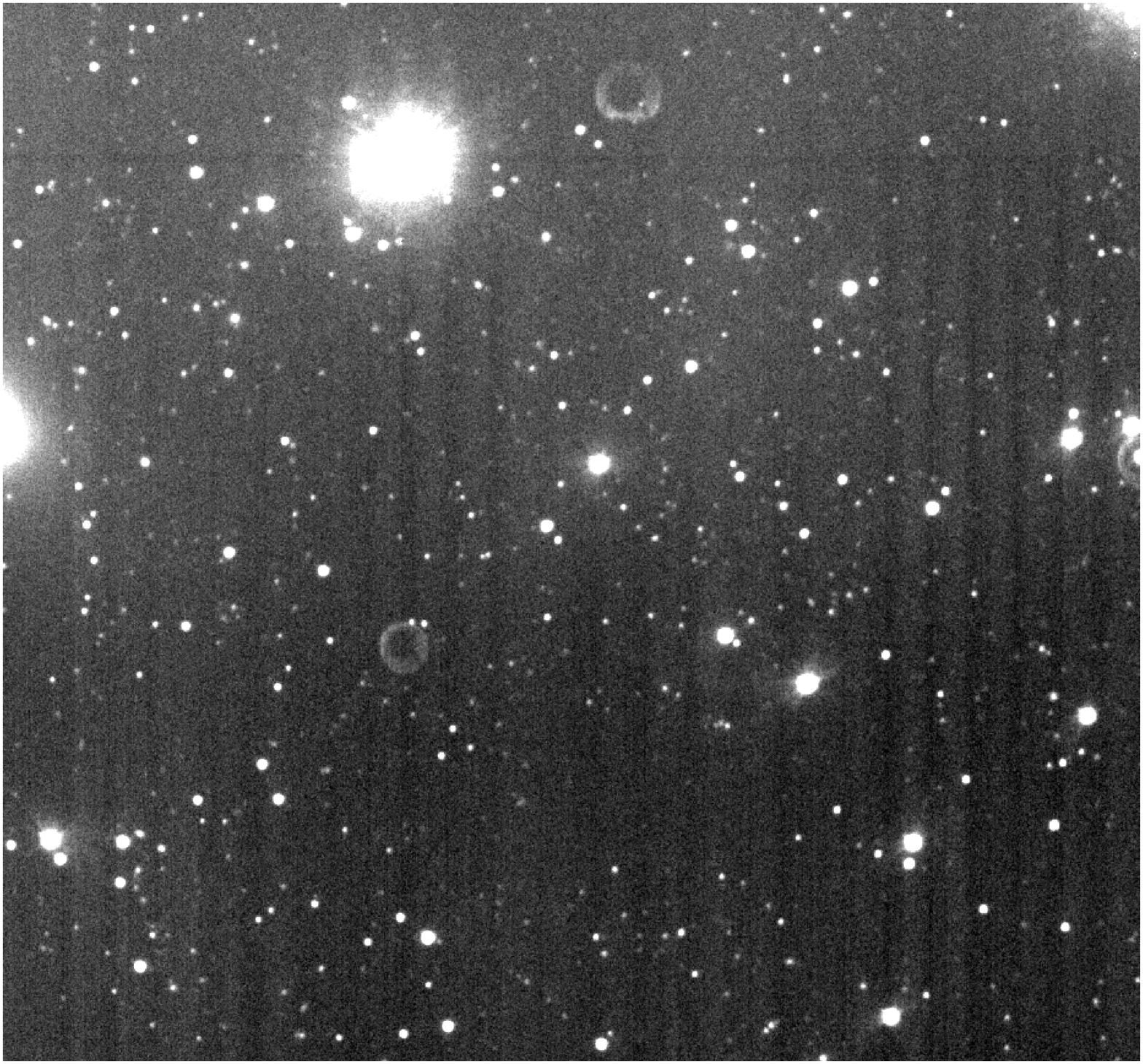} \\
        {\scriptsize Set3} & {\scriptsize Set4} \\
        \includegraphics[width=0.4\hsize]{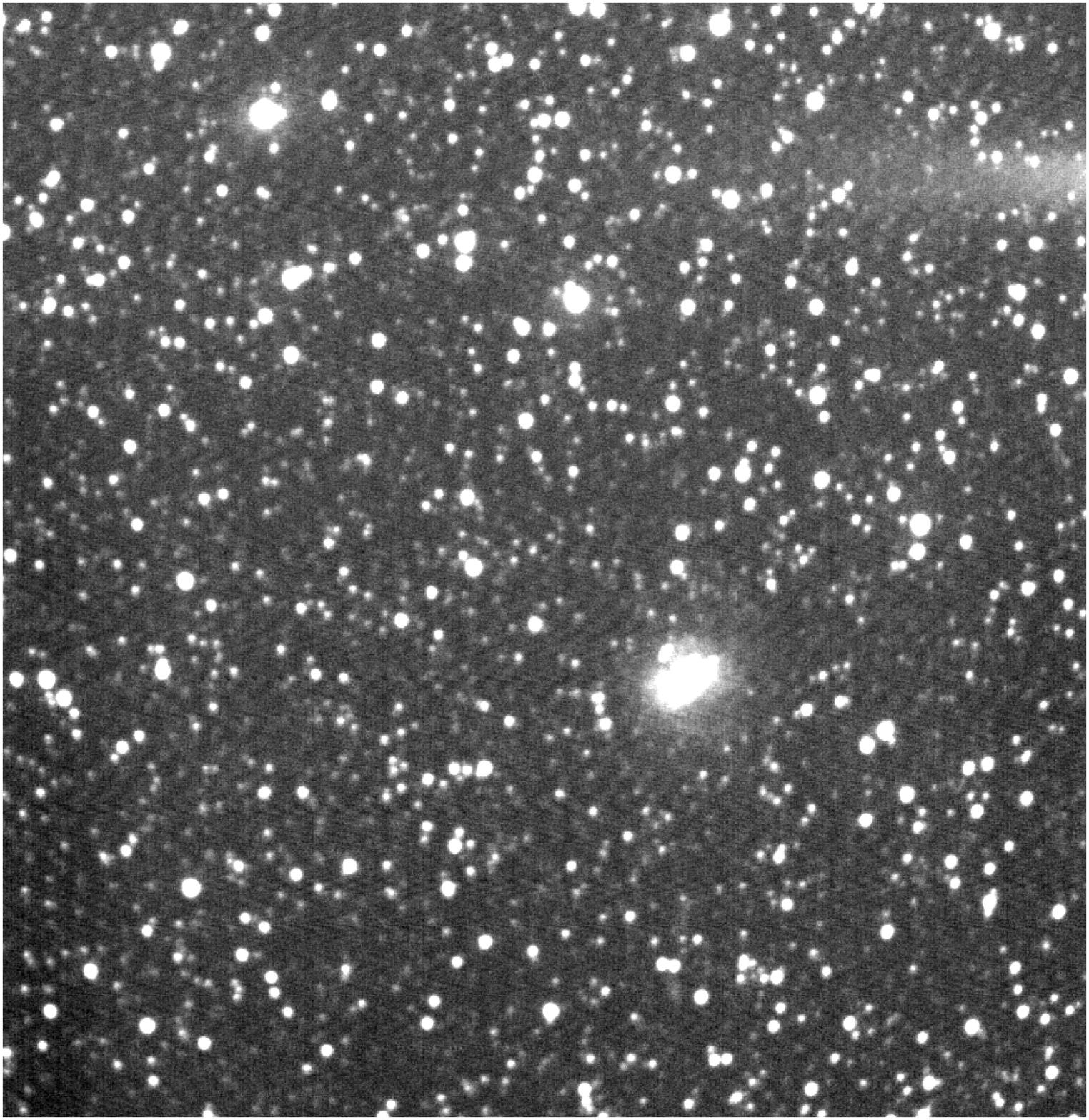} & \includegraphics[width=0.4\hsize]{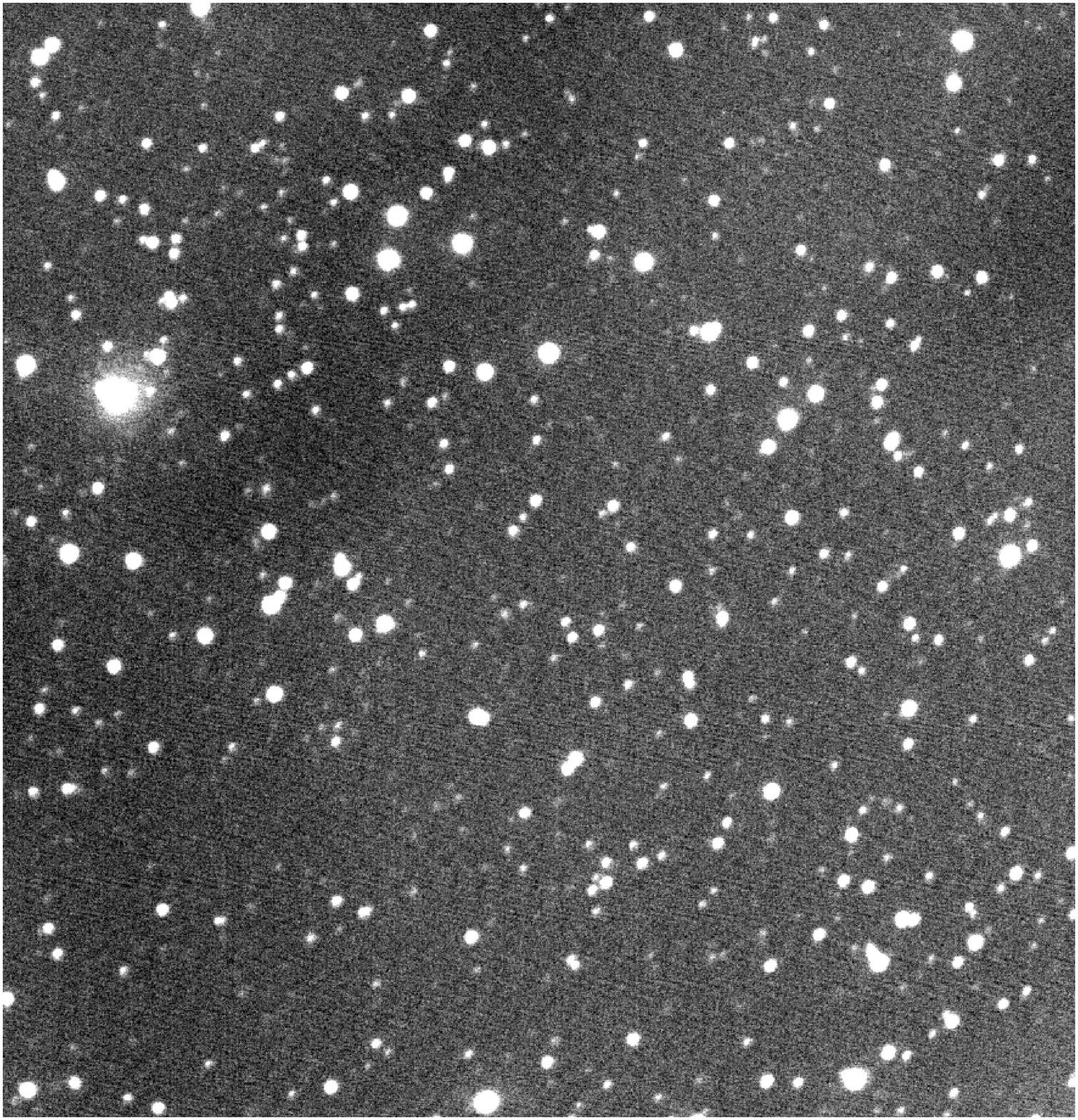}
    \end{tabular}
    \end{center}
    \caption{Co-added images of data sets used for verification of co-adding consistency}
    \label{fig:combine}
\end{figure}

For the data set processed up to the alignment, we generated a co-added image by a function of the new pipeline, and calculated $\Delta\hat{I}$ using a image co-added by IRAF-\texttt{imcombine} as the reference.
The input images are acquired at MITSuME-Akeno on May 14, 2018. Information about the input-image sets are given in table \ref{tab:combine}. The co-added images are shown in figure \ref{fig:combine}.

\begin{table}
  \tbl{
    Information of data-sets used for verification of co-adding consistency}{
    \begin{tabular}{lrr}\hline
    Label\footnotemark[1] & Number of Images & Shape\footnotemark[2] (X$\times$Y) [pix]\\\hline
    Set1 & 35 & $910\times944$\\
    Set2 & 81 & $928\times864$\\
    Set3 & 97 & $911\times938$\\
    Set4 & 44 & $811\times916$\\\hline
    \end{tabular}
    }\label{tab:combine}
\begin{tabnote}
\footnotemark[1] An appropriate name to identify the data-set\\
\footnotemark[2] The reason why the shape of the image is different from that before the alignment is that an area which does not overlap with another image is cut out during the alignment.\\
\footnotemark[$*$] For all data-set, the numerical format is a 32-bit float.
\end{tabnote}
\end{table}

\begin{figure}
    \begin{center}
        \includegraphics[width=\hsize]{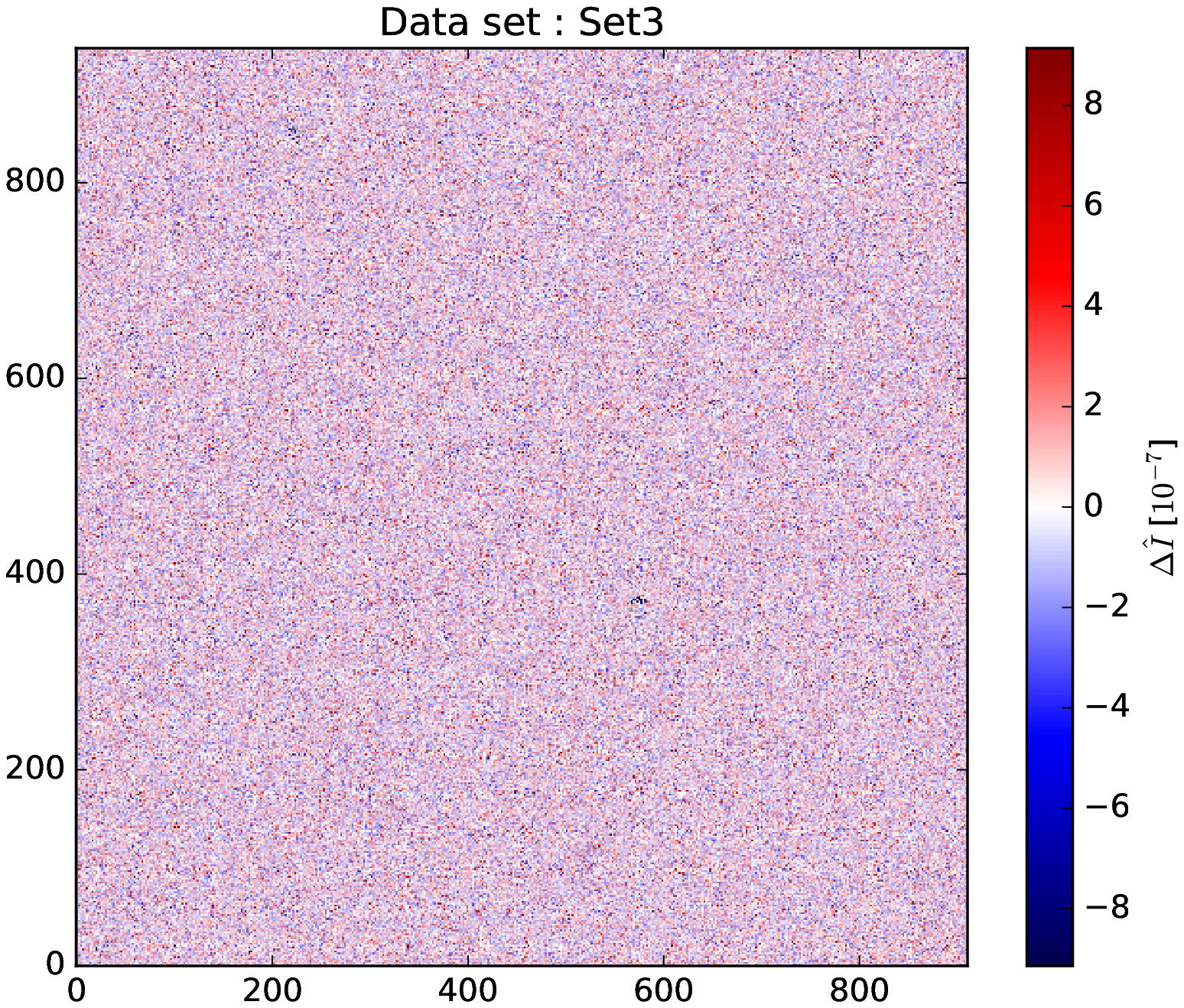}   
    \end{center}
    \caption{An image in which each pixel is appropriately colored with $\Delta\hat{I}$. Data set is Set3. No particular pattern appears.}\label{fig:show_comb}
\end{figure}

\begin{figure}
    \begin{center}
        \includegraphics[width=\hsize]{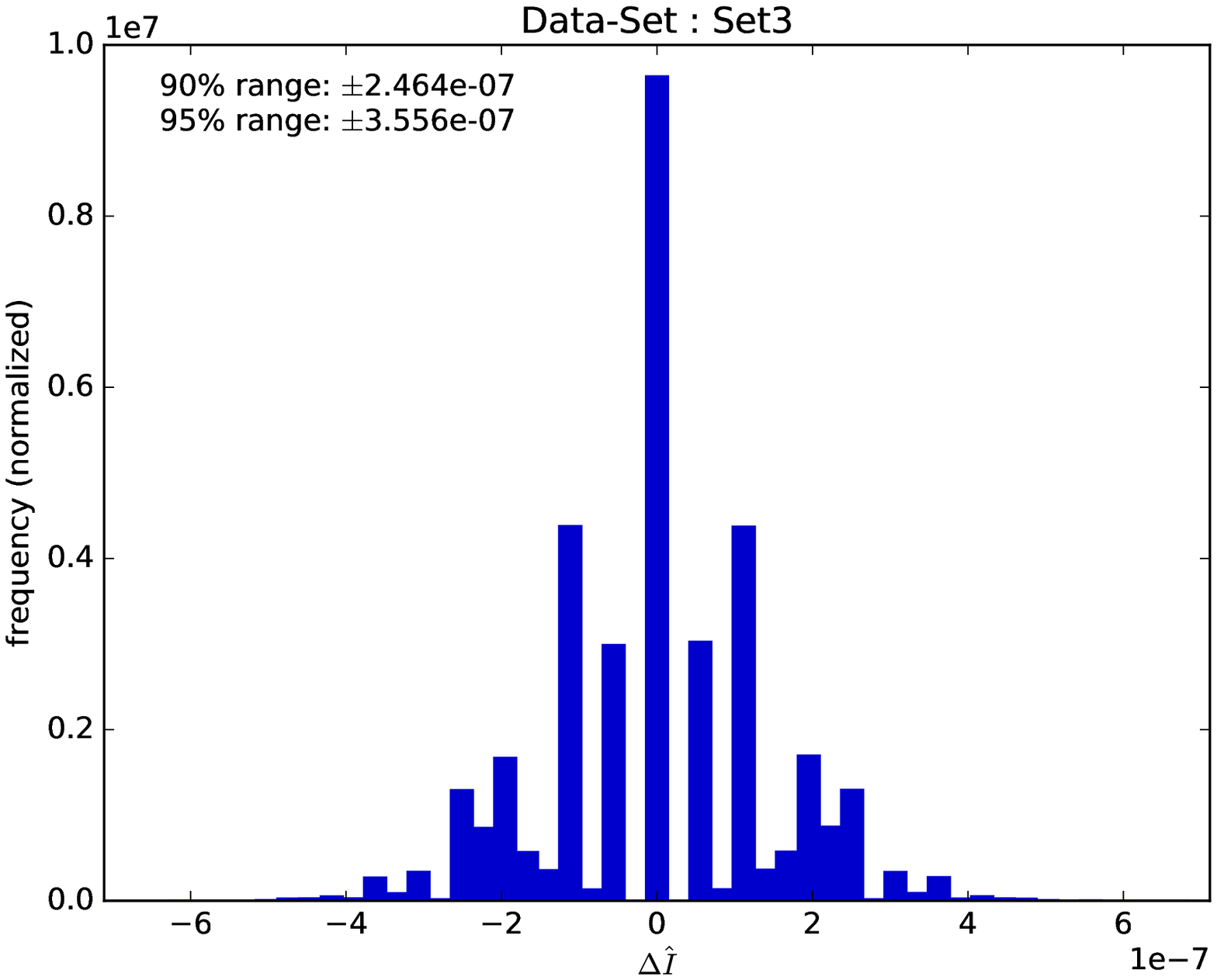}        
    \end{center}
    \caption{A histogram of $\Delta\hat{I}$. Data set is Set3. ``90(95)\% range'' is the range in which 90(95)\% of Diff of all pixels fall. The value of Diff is discrete, and is about $10^{-7}$.}\label{fig:hist_comb}
\end{figure}

Typical results are shown in figure \ref{fig:show_comb} and figure \ref{fig:hist_comb}.
Figure \ref{fig:show_comb} is a image in which each pixel is appropriately colored with $\Delta\hat{I}$, and figure \ref{fig:hist_comb} is a histogram of $\Delta\hat{I}$.
In figure \ref{fig:show_comb}, no particular pattern is seen.
As can be seen from figure \ref{fig:hist_comb}, $\Delta\hat{I}$ is discrete, and is about $10^{-7}$.
32-bit floating point conforming to IEEE754 is assigned 1 bit for a sign, 8 bits for an exponent, and 23 bits for a significand.
In the case of a non-zero number, the maximum digit of the significand is always 1.
So, the significand is regarded as having 24 bits by assuming that there is always 1 in the 24th digit \citep{ieee754}.
Therefore, the maximum number of decimal digits that can be represented by 32-bit floating point is $\log_{10}{2^{24}}\approx7$.
Since the last digit can easily fluctuate depending on the order of calculations, the maximum number of significant digits is about 6, and this is comparable to our result.
Therefore, it can be said that the co-adding function of the new pipeline reproduces the same function as IRAF-\texttt{imcombine} within the precision of 32-bit floating point.

\subsection{Execution Speed}\label{sec:speed}
The procedure for measuring the execution time is described below.
Using a set of images that have not been reduced as input, we ran the pipeline 5 times while displaying the execution time with the Unix command \texttt{time}.
We considered an average of the five output values as the execution time in this condition, and the standard deviation as the error.
We define the number of inputs for the three bands as $n$ and the execution time as $t$.

\subsubsection{Comparison with the current pipeline}\label{sec:speed-comparison}
We measured $t$ while changing $n$ by 30 (10 per band) for the current pipeline, the new one, and the new one running entirely on the CPU without using the GPU.
The third is a code in which processing performed by CuPy is rewritten with NumPy to separate the effect of performing processing on the GPU from nonusage of IRAF/PyRAF.
Thereafter, we refer to the current pipeline as ``Current'', the new pipeline as ``New(GPU)'', and the CPU version of the new pipeline as ``New(CPU)''.
In order to evaluate the difference of the calculation time between the CPU and GPU as purely as possible, we also measured the breakdown of $t$ for New(CPU) and New(GPU).
This was performed by acquiring the UNIX time in each processing step using the Python module \texttt{time}\footnote{Python build-in module that provides time-related functions}.

We used a set of images acquired with MITSuME-Akeno at August 22, 2018.
188 images per band were captured.
In order to make a nice round number, the first input was 180 images per band (i.e. $n=540$).

\begin{figure}
    \begin{center}
        \includegraphics[width=\hsize]{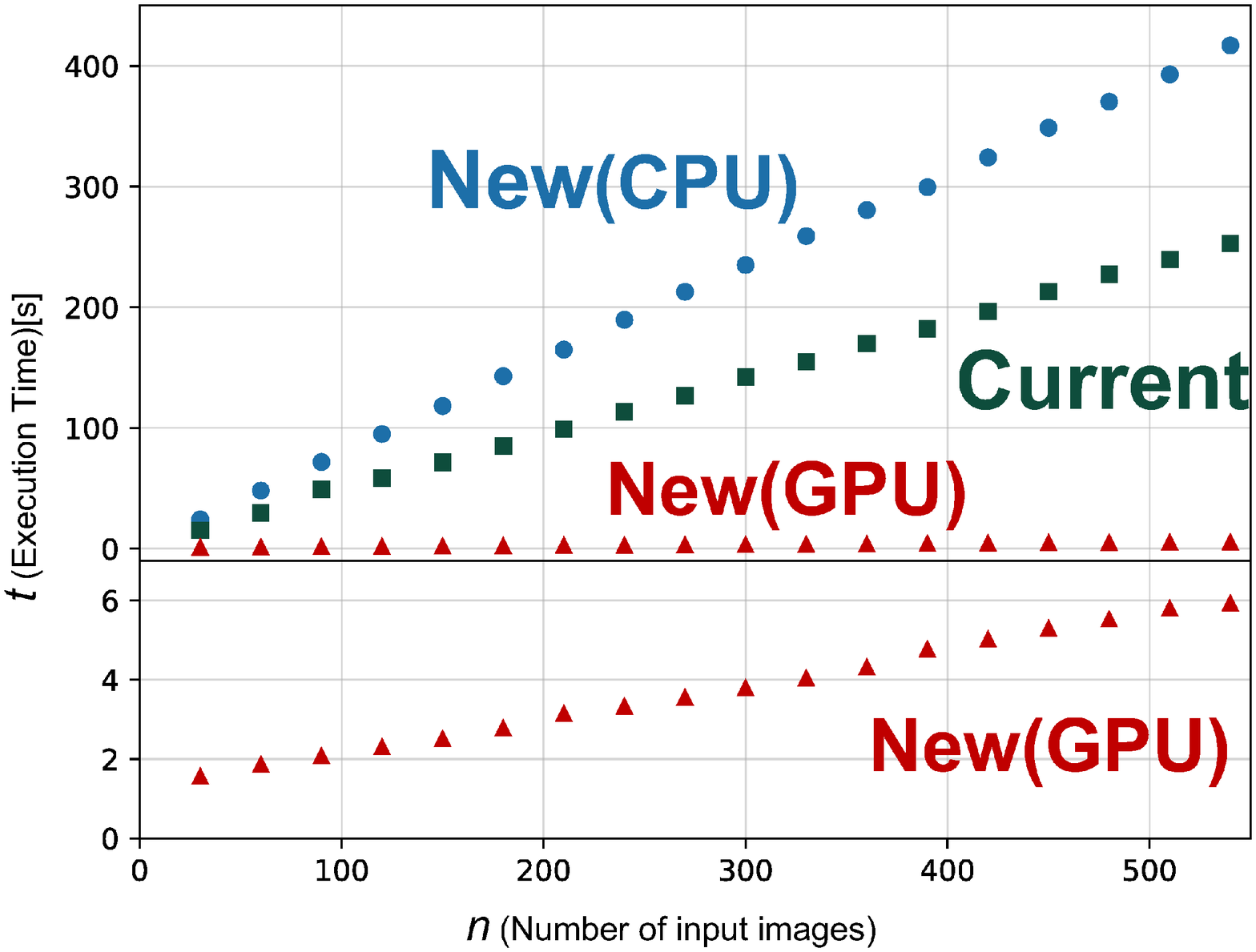}   
    \end{center}
    \caption{
    Plot of $n$ and $t$ for each pipeline.
    Errors are not drawn because they are comparable to the marker size.
    It is hard to see the plot of New(GPU) in the above figure, so the scale changed figure is shown below.
    For $n\le540$, $t$ of each pipeline increases almost linearly with $n$, and New(GPU) is significantly faster than Current and New(CPU).
    New(CPU) is slower than Current because our implementation of the bicubic-spline interpolation for image alignment is less efficient than IRAF-\texttt{imshift}.}\label{fig:time_linear}
\end{figure}
Figure \ref{fig:time_linear} is a plot of $n$ and $t$ for each pipeline.
It shows that $t$ increases linearly with respect to $n$ for $n\le540$.
This is because both the file I/O amount and the calculation amount are approximately proportional to the data volume.
\begin{figure}
    \begin{center}
        \includegraphics[width=\hsize]{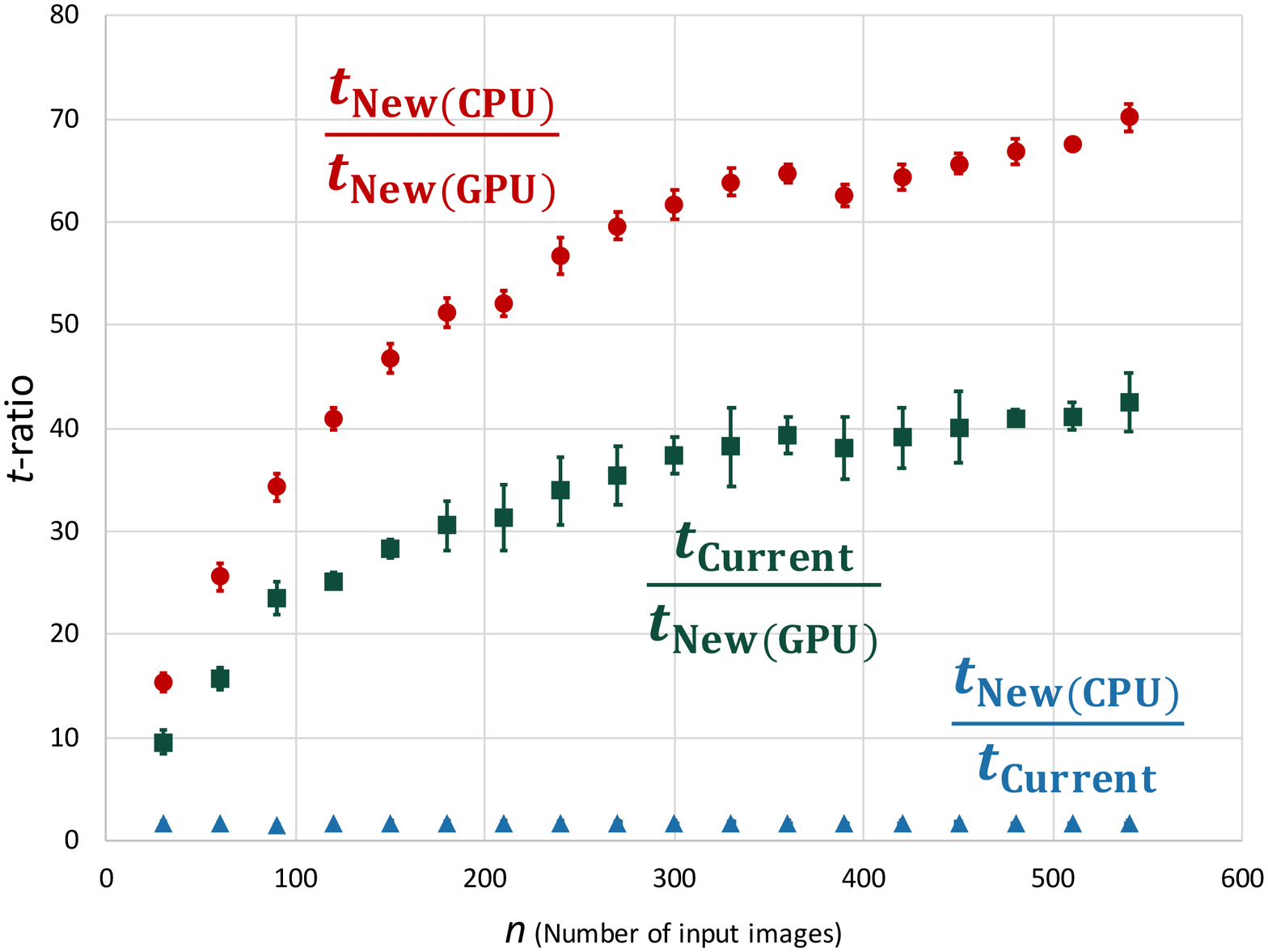}   
    \end{center}
    \caption{Plot of $n$ and the ratio of $t$. When $n = 540$, the ratio of $t$ between Current and New (GPU) is 42. That is, New(GPU) runs 42-times faster.}\label{fig:time_ratio}
\end{figure}
Figure \ref{fig:time_ratio} is a plot of $n$ and $t$-ratio for each pipeline.
The ratio differs depending on $n$ because $t$ has a constant component for $n$ (e.g. starting up the interpreter).
When $n = 540$, the ratio of $t$ between Current and New(GPU) is 42. That is, New(GPU) runs 42 times faster.
\begin{table}
  \tbl{Breakdown of $t$}{
    \begin{tabular}{llll}\hline
    ($n=540$)& New(GPU) [s] & New(CPU) [s] & Ratio\footnotemark[1]\\\hline
    FITS I/O\footnotemark[2] & 3.4 & 2.7 & 0.81\\
    WCS operation & 0.96 & 0.94 & 0.97\\
    Bias,Dark,Flat & $1.0\times10^{-3}$ & 0.74 & 720\\
    Bad-pixel mask & 0.36 & 10.55 & 29\\
    Alignment & 0.48 & $4.0\times10^2$ & 830\\
    Combining & 0.14 & 12 & 91\\\hline
    \end{tabular}
    }\label{tab:time_items}
\begin{tabnote}
\footnotemark[1] New(CPU)/New(GPU)\\
\footnotemark[2] Including RAM-VRAM transfer time\\
\footnotemark[$*$] Interpreter startup and module initialization time are not shown
\end{tabnote}
\end{table}
Table \ref{tab:time_items} shows the breakdown of $t$ for New(CPU) and New(GPU).
Since FITS I/O and WCS operation do not use the GPU for calculation even in New(GPU), the results are comparable between New(GPU) and New(CPU).
On the other hand, the other processing clearly shows differences among the calculation speed of GPU and CPU, New(GPU) is about 10-1000 times faster than New(CPU).
In particular, since the calculational cost of bicubic-spline interpolation is very large, New(CPU) spends a very long time on alignment.
Therefore, if a less costly algorithm such as nearest-neighbor or bilinear interpolation is used, $t$ is expected to be significantly reduced.
In our environment, NumPy uses only a single CPU core.
However, depending on how NumPy is built, multiple cores can be used, and it is expected to run faster.
We note that New(CPU) is slower than Current by factor of 2.
This is probably caused by the different implementations of their bicubic-spline interpolations in the image alignment.
In the bicubic-spline interpolation, there is a process of solving simultaneous linear equations for each column or row of the image.
The number of unknowns of each simultaneous equations is about the same as $N$, the number of pixels in the each column or row (e.g. $N=1020$ for MITSuME data), and those simultaneous equations can be expressed as a tridiagonal matrix.
The cost of the solution optimized for the tridiagonal matrix is $O(N^2)$, and IRAF-\texttt{imshift} probably uses such a solution.
The new pipeline solves those equations using the inverse matrix which results in the cost of $O(N^3)$.

\subsubsection{Application to various data}\label{sec:speed-application}
\begin{figure}
    \begin{center}
        \includegraphics[width=\hsize]{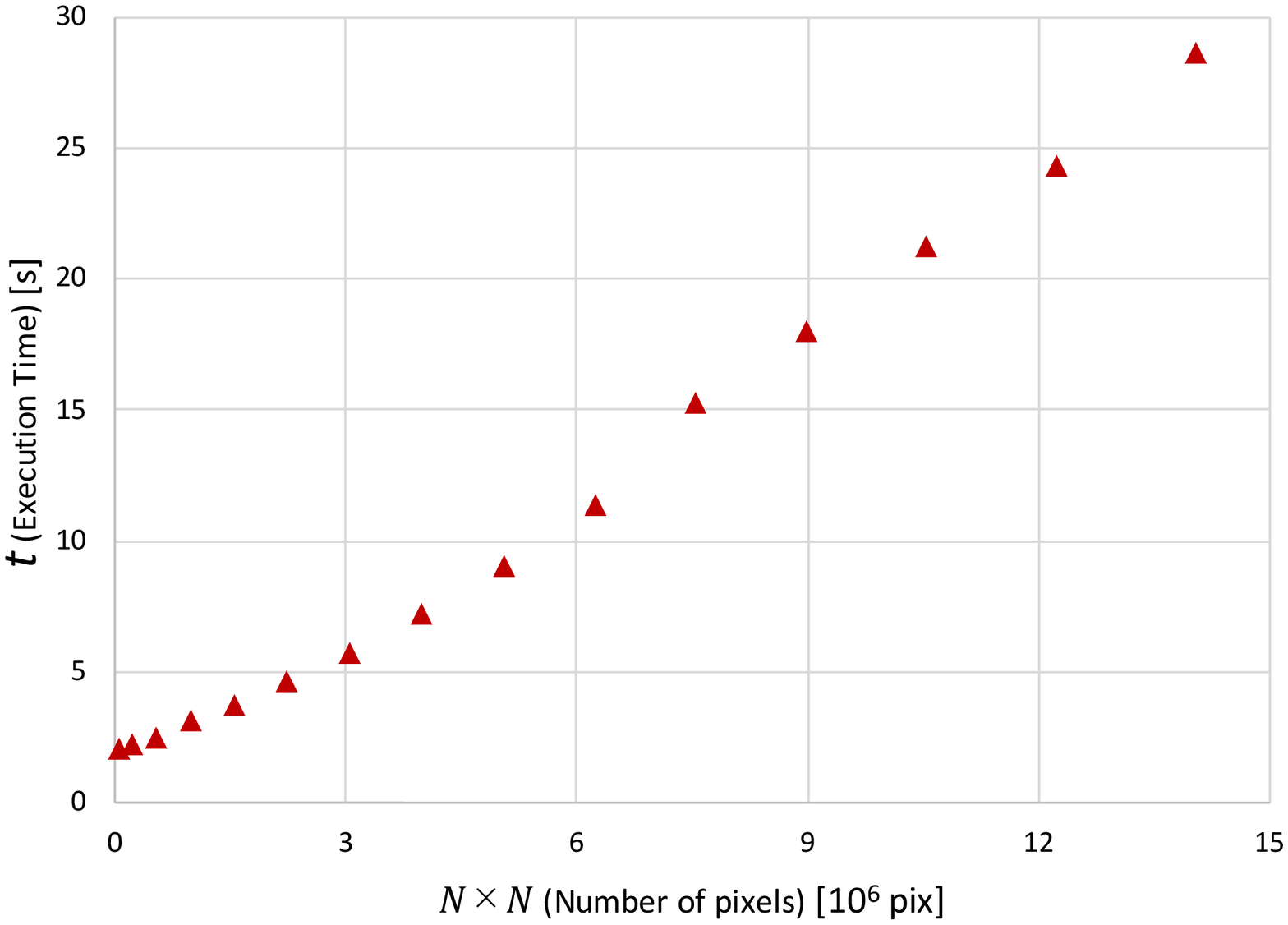}   
    \end{center}
    \caption{Plot of $N\times N$ and $t$. $t$ has a downward convex change with respect to $N^2$. It seems that linear fitting can be performed up to $N^2\le3\times10^6$, but the same fitting cannot be applied to all regions.}\label{fig:nop-time}
\end{figure}

We also evaluated the execution time for the various image sizes, assuming the application of this pipeline to observational data other than MITSuME.
We generated a set of dummy images in which random values were assigned to $N\times N$ pixels, and measured the execution time $t$ while changing $N$ as long as the VRAM is sufficient.
The number of images was fixed at 100 per band (i.e. $n=300$). The result is shown in figure \ref{fig:nop-time}.

\begin{figure}
    \begin{center}
        \includegraphics[width=\hsize]{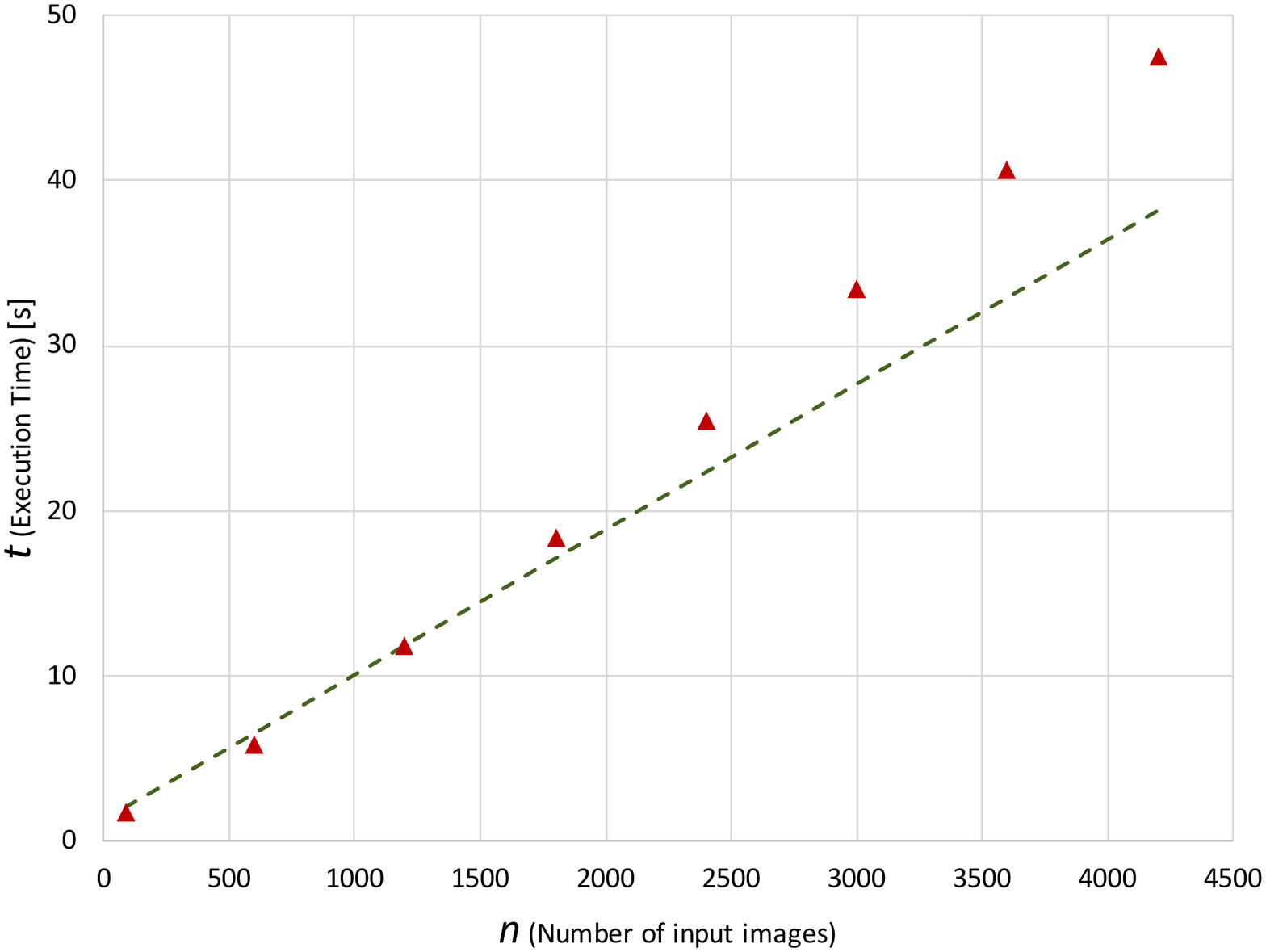}   
    \end{center}
    \caption{Plot of $n$ and $t$. This is the same condition of New(GPU) in figure \ref{fig:time_linear}, except that the input is dummy data.
    The dashed line is the model obtained by linear-fitting the plot of New(GPU) in figure \ref{fig:time_linear}. At $n>1000$, the plot deviates significantly from the line and has a convex downward change.}\label{fig:noi-time}
\end{figure}

Assuming that both the file-reading time and the calculation time are basically proportional to the data volume, $t$ is considered to change linearly with respect to $N^2$.
However, in figure \ref{fig:nop-time}, $t$ has a convex downward change with respect to $N^2$.
The first conceivable cause is the calculation time of bicubic-spline interpolation.
As mentioned in section \ref{sec:speed-comparison}, our implementation of the bicubic-spline interpolation has a calculational cost of $O(N^3)$.
Therefore, this effect does not depend on hardware.
Figure \ref{fig:noi-time} shows the result of performing the same measurement while fixing $N=1020$ and changing $n$ as long as the VRAM is sufficient.
This is the same condition as the measurement performed in section \ref{sec:speed-comparison}, except that the input data is dummy data.
When $n$ is large, $t$ has a downward convex change respect to $n$.
The details have not been clarified yet, but we believe that a non-linear effect about the data volume is affecting the execution time.
What we currently suspect is the time required for the memory allocation and releasing.
However, we could not reproduce that effect by simply allocating or releasing the memory.

\section{Summary}
To speed up the image reduction processing of MITSuME data, and to establish set of a low-cost, high-speed image-processing methods, we developed a GPU-accelerated image-reduction pipeline.
This pipeline uses CuPy, a Python package for GPGPU, to perform image processing on the GPU.
Because the library does not provide a corresponding function, there is some processing which was coded by ourselves.
However, this pipeline reproduced the functions of the current pipeline with a relative difference of $\lesssim10^{-5}$.
Furthermore, the new pipeline is 42 times faster than the current pipeline when processing 540 images.

Currently, we are performing real-time image reduction of MITSuME observational data with GPU-accelerated pipeline, and using it to check results of prompt follow-up observation of GRBs.
At present, the reduced images of this pipeline have been used in 6 reports to GRB Coordinates Network Circular Service \citep{gcnc1, gcnc2, gcnc3, gcnc4, gcnc5, gcnc6}.

We have developed a Python package based on our new pipeline and released it on GitHub\footnote{https://github.com/MNiwano/Eclaire}.
Thus, GPU-accelerated image reduction can be performed with on data from other observatories. 



\begin{ack}
This work was supported by JSPS KAKENHI Grant Numbers JP16K13783.
This work was also supported by Optical and Near-Infrared Astronomy Inter-University Cooperation Program of the MEXT of Japan, and by JSPS and NSF under the JSPS–NSF Partnerships for International Research and Education.
\end{ack}





\end{document}